\documentclass[%
reprint,nosuperscriptaddress
amsmath,amssymb,nofootinbib,
aps,
prd,longbibliography,
]{revtex4-2}
\usepackage{graphicx}
\usepackage{dcolumn}
\usepackage{bm}
\usepackage[colorlinks=true,linkcolor=blue,citecolor=blue,urlcolor=blue]{hyperref}
\usepackage{booktabs}
\usepackage{amsmath}
\usepackage{multirow}
\allowdisplaybreaks[1]  
\usepackage{siunitx} 

\def \mS {\mathcal S}
\begin{document}
	
\title{Tidal Love numbers for regular black holes}

\newcommand{\SOUTHCUT}{\vspace{0.5em} School of Physics and Optoelectronics, South China University of Technology, \\ Guangzhou 510641, People's Republic of China}

\author{ Rui Wang}
\author{Qi-Long Shi}
\author{Wei Xiong}
\author{Peng-Cheng Li}
\email{pchli2021@scut.edu.cn}

\affiliation{\SOUTHCUT}

\date{\today}

	\begin{abstract}
	Tidal Love numbers (TLNs) characterize the response of compact objects to external tidal fields and vanish for classical Schwarzschild and Kerr black holes in general relativity. Nonvanishing TLNs therefore provide a potential observational window into beyond-classical physics. In this work, we present a unified and fully analytic study of the TLNs of three representative classes of regular black holes---the Bardeen black hole, the black hole with sub-Planckian curvature, and the black hole arising in asymptotically safe gravity---under scalar, vector, and axial gravitational perturbations. Employing a Green's function method combined with systematic perturbative expansions, we show that TLNs of regular black holes are generically nonzero and exhibit strong model and mode dependence. In many cases, higher-order corrections develop logarithmic scale dependence, closely resembling renormalization-group running in quantum field theory and revealing a scale-dependent tidal response absent in classical black holes. Our analysis demonstrates that the internal structure of regular black holes, including de Sitter or Minkowski cores and quantum-gravity-inspired modifications, leaves distinct fingerprints in their tidal properties. These results provide a comparative theoretical benchmark for assessing regular black-hole models and establish a basis for future phenomenological and observational studies with gravitational-wave detectors.
	\end{abstract}
	
	\maketitle

\section{Introduction}
General relativity (GR) predicts the existence of both gravitational waves (GWs) and black holes (BHs). The direct detection of GWs from binary BH mergers has provided strong observational confirmation of these predictions~\cite{LIGOScientific:2016aoc}. During the early inspiral phase of a binary coalescence, the orbital separation gradually decreases, and tidal interactions---if present---become increasingly important. Such tidal interactions are characterized by the \emph{tidal Love numbers} (TLNs), which quantify the response of a compact object to an external tidal field by relating the induced multipole moments to the applied tidal potential. The concept of Love numbers was first introduced by A.~E.~H.~Love in 1909~\cite{love1909yielding} to describe the Earth's deformation under lunar and solar tides. In relativistic astrophysics, TLNs have become essential observables linking the internal structure of compact objects to GW signals, and are now used to constrain neutron star equations of state~\cite{Flanagan:2007ix,LIGOScientific:2018cki}.

In GR, a remarkable property of four-dimensional, asymptotically flat BHs is the vanishing of their TLNs. This was first shown for Schwarzschild BHs ~\cite{Fang:2005qq,Binnington:2009bb,Damour:2009vw,Kol:2011vg, Gurlebeck:2015xpa,Hui:2020xxx}; the same result was then shown to apply to Reissner-Nordström~\cite{Cardoso:2017cfl,Pereniguez:2021xcj,Rai:2024lho} (however, fermionic tidal perturbations may induce norzero TLNs \cite{Chakraborty:2025zyb,Pang:2025myy}), Kerr~\cite{Landry:2015zfa,Pani:2015nua,Poisson:2014gka,LeTiec:2020bos,LeTiec:2020spy,Chia:2020yla,Charalambous:2021kcz,Charalambous:2021mea,Ivanov:2022qqt} and Kerr-Newman~\cite{Ma:2024few} BHs.  However, deviations from this prediction can arise in modified gravity theories~\cite{Cardoso:2018ptl,Cardoso:2017cfl,DeLuca:2022tkm,Barura:2024uog,Bhattacharyya:2025slf,Katagiri:2024fpn} or when BHs are surrounded by matter~\cite{Cardoso:2019upw,DeLuca:2022xlz,Cannizzaro:2024fpz,Chakraborty:2024gcr}. On the other hand, exotic compact objects, including BH mimickers, wormholes, and ultracompact stars, typically possess nonzero TLNs~\cite{Cardoso:2017cfl,Cardoso:2019rvt,Chirenti:2020bas,Collier:2022cpr,Nair:2022xfm}. The existence of nonvanishing TLNs in such cases could therefore serve as a smoking gun for deviations from GR ~\cite{Cardoso:2017cfl, Maselli:2017cmm,Datta:2021hvm,Chia:2023tle,Piovano:2022ojl,Zi:2023pvl,Andres-Carcasona:2025bni}

Regular BHs (RBHs)---nonsingular solutions that avoid curvature divergences at the center---have been extensively explored as phenomenological models of quantum-corrected spacetimes~\cite{Lan:2023cvz,Bambi:2023try}. Examples include the Bardeen BH, which introduces a de~Sitter-like core to remove singularities~\cite{bardeen1968, Ayon-Beato:2000mjt}; the  BH with sub-Planckian curvature, which maintains finite curvature for all masses~\cite{Ling:2021olm,Xiang:2013sza}; and BHs arising in asymptotically safe gravity (ASG)~\cite{Bonanno:2000ep, Platania:2023srt,Bonanno:2023rzk,Spina:2025wxb}, which embody key aspects of quantum-gravity renormalization effects. Despite increasing interest, a systematic investigation of TLNs in these RBH scenarios remains incomplete \cite{Motaharfar:2025typ,Coviello:2025pla,Motaharfar:2025ihv,Liu:2025iby}. Understanding how their internal geometries and quantum-inspired corrections influence tidal responses is therefore crucial for connecting theoretical models to future GW observations.

In this work, we perform a comprehensive analysis of the TLNs of three representative classes of RBHs: the Bardeen BH \cite{bardeen1968}, the BH with sub-Planckian curvature~\cite{Ling:2021olm}, and the BH formed through gravitational dynamics during collapse within ASG \cite{Bonanno:2023rzk}, under scalar, vector and axial gravitational perturbations. To obtain these results analytically, we adopt a Green’s function technique developed in Ref. \cite{Barura:2024uog}. Our goal is not to focus on formalism development, but to use a unified framework to clarify how different regular BH geometries modify the tidal response under scalar, vector, and axial gravitational perturbations. We show that the TLNs of these regular BHs are generically nonvanishing, with characteristic dependence on the perturbation sector, multipole index, and deviation parameters. In many cases, higher-order corrections also exhibit logarithmic scale dependence, which can be interpreted in terms of classical renormalization-group (RG) running. Taken together, these results identify a set of comparative tidal signatures that can serve as useful theoretical benchmarks for future studies of RBHs and related compact-object phenomenology. A quantitative assessment of whether these model-dependent tidal signatures can be resolved by specific gravitational-wave observations requires dedicated waveform-based analyses and lies beyond the scope of the present paper.


The structure of this paper is as follows.
Section~\ref{backgroung} introduces the background metrics of the three representative RBH models. 
Sec.~\ref{sectionIII} presents the formalism and Green’s function technique used to compute the TLNs. 
Section~\ref{sectionIV} discusses the obtained results for scalar, vector, and axial perturbations, highlighting their physical implications. 
Finally, Sec.~\ref{sectionV} concludes with a summary of our findings and an outlook for future GW tests of RBH models. We use geometric units $G =c=1$ everywhere.

\section{Background OF regular BLACK HOLES}\label{backgroung}
In this section, we briefly review the three aforementioned models of RBHs, focusing on their metric representations and fundamental properties. The metrics for all three can be uniformly expressed in Schwarzschild coordinates as
\begin{equation}\label{metricref}
	ds^2 = - f(r) \, dt^2 + \frac{dr^2}{f(r)} + r^2 d\Omega^2,
\end{equation}
where $d\Omega^2=d\theta^2+\sin^2\theta d\phi^2$ is the metric on the unit two-sphere. The distinguishing feature of each specific RBH model is then given by its unique metric function $f(r)$.

\subsection{Bardeen black holes}

The Bardeen BH stands as the first RBH in GR, which was initially introduced as a phenomenological model \cite{bardeen1968}. Ayón-Beato and García provided a physical interpretation of the Bardeen BH by showing that it can be sourced by a magnetic monopole within a nonlinear electrodynamics framework~\cite{Ayon-Beato:2000mjt}. In addition, quantum-gravity corrections derived from the generalized uncertainty principle have also been suggested as an alternative mechanism for generating the Bardeen BH~\cite{Maluf:2018ksj}. 

The metric function of the Bardeen BH takes the form
\begin{equation}\label{eq:metricfuncs}
	f(r)=1-\frac{2 M r^{2}}{(r^{2}+q^{2})^{3/2}},
\end{equation}
where $M$ is the mass of the BH and the parameter $q$ acts as a regularization parameter.  The solution continuously reduces to the Schwarzschild spacetime in the limit $q \to 0$.  As shown in Ref. ~\cite{Ayon-Beato:2000mjt}, the spacetime has horizons only when
$q\leq\tfrac{4M}{3\sqrt{3}} \approx0.77M$. For larger values of $q$, the spacetime has no horizons.
A notable feature of the Bardeen BH is that the central singularity is replaced by a regular core, which behaves effectively as a de Sitter region.
This becomes clear from the small-$r$ expansion, where  $f(r) \to1- 2M r^{2}/q^3$, as $r\to0$. Dymnikova \cite{Dymnikova:1992ux} showed that, within GR, any static spherically symmetric BH with a regular center and matter obeying the weak energy condition must approach a de Sitter core at the center. Well-known examples include the Hayward BH~\cite{Hayward:2005gi} and the Frolov BH~\cite{Frolov:2016pav}, in addition to the Bardeen BH.

To study the gravitational perturbations of all three RBHs within a unified framework, we adopt the perspective that the regularity of the Bardeen BH arises from effective quantum corrections rather than from nonlinear electrodynamics, in analogy with the mechanisms underlying the other two RBH models.

\subsection{Regular black holes with sub-Planckian curvature}

Alternative to the Bardeen BH, Ref.~\cite{Xiang:2013sza} analyzed the general conditions required for obtaining a finite Kretschmann scalar and  proposed a RBH solution featuring a Minkowski core. In this scenario, as $r \to 0$, the metric functions satisfy  $-g_{tt} =g_{rr}\to 1$, so that the central singularity is replaced with a flat Minkowski spacetime instead of a de Sitter core.
However, Ling and Wu~\cite{Ling:2021olm} pointed out that in the RBH proposed in Ref.~\cite{Xiang:2013sza}, the maximum value of the Kretschmann scalar depends on the BH mass. For a fixed short-distance regulator, the peak curvature scales as $M^2$, meaning that it can grow arbitrarily large and exceed the Planck scale $M_p^4$ for sufficiently massive BHs. This behavior conflicts with the expectation from quantum gravity that spacetime curvature should remain bounded. 

Motivated by this issue, Ling and Wu~\cite{Ling:2021olm} constructed a  class of RBHs with a Minkowski core, based on the generalized uncertainty principle in curved spacetime, in which the Kretschmann scalar remains sub-Planckian for any BH mass. At large radii, this metric can reproduce the asymptotic behavior of Bardeen, Hayward, or Frolov BHs, depending on the choice of the potential form. The metric function of the RBH with a Minkowski core and bounded curvature is
\begin{equation}
	f(r) = 1 + 2 \psi(r) = 1 - \frac{2 M}{r} \, e^{-\alpha_0 M^x / r^c},
\end{equation}
where $\psi(r)$ denotes the modified Newton potential, $M$ is the mass of the BH, and $\alpha_0$, $x$ and $c$ are dimensionless parameters. The parameters $\alpha_0$ characterizes the deviation from the Newtonian potential and encodes quantum-gravity corrections. To ensure the existence of a horizon and keep the curvature below the Planck scale, the parameters must satisfy  $c\geq x\geq c/3$ and $c\geq2$. Typically, when $\alpha_0=0$, the metric reduces to the Schwarzschild solution. For  $x = 2/3$ and $c = 2$, the metric matches the large-$r$ asymptotics of the Bardeen BH, while choosing  \(x=1\) and \(c=3\) yields the Hayward-like asymptotic behavior. In this work, we focus on the case \(x=1\) and \(c=3\).

\subsection{ A regular black hole in asymptotically safe gravity}

The asymptotic safety scenario provides a compelling candidate framework for quantum gravity, whose defining feature is the existence of a non-Gaussian ultraviolet fixed point of the RG flow. In this approach, the scale dependence of Newton’s constant is governed by RG trajectories that approach the UV fixed point at high energies, thereby preventing divergences in the gravitational coupling and ensuring the predictive power of the theory \cite{Niedermaier:2006ns}.

Recently, Bonanno {\em et al}. \cite{Bonanno:2023rzk} extended the proposal of Markov and Mukhanov \cite{Markov:1985py} and investigated the gravitational collapse of dust (a pressureless ideal fluid) within the framework of ASG. Guided by the Reuter fixed point \cite{Bonanno:2000ep}, the authors introduced a multiplicative coupling between the matter sector and gravity, which effectively implements the weakening of the gravitational interaction at high energies. This modification leads to an effective Lagrangian that yields a regular interior geometry, which can be smoothly matched to a static and asymptotically flat exterior spacetime. The matching is uniquely fixed by requiring that both the induced metric and the extrinsic curvature remain continuous at the boundary of the collapsing matter distribution when approached from either side.

The resulting static, spherically symmetric exterior line element takes the form
\begin{equation}
	f(r) = 1 - \frac{r^2}{3 \, \xi} \, \ln\left( 1 + \frac{6 M \, \xi}{r^3} \right),
\end{equation}
where $M$ denotes the BH mass and $\xi$ is a scale parameter characterizing the deviation from the Schwarzschild solution. The precise value of $\xi$ is currently unknown and should be constrained observationally. In the limit $\xi \to 0$, the Schwarzschild geometry is recovered. 
A numerical analysis of the horizon condition $f(r_h)=f'(r_h) = 0$, 
reveals a critical value $\xi_{\rm cr}\simeq 0.4565 \, M^2$ with $r_h \simeq 1.2516 \, M$. 
For $0 < \xi < \xi_{\rm cr}$, the metric exhibits two event horizons, an inner and an outer one. 
The two horizons merge into a single degenerate surface for $\xi = \xi_{\rm cr}$, corresponding to an extremal BH. For $\xi > \xi_{\rm cr}$, no horizon forms.

\section{Tidal Love Numbers Using the Green's Function Technique}\label{sectionIII}
In this section, we briefly review the definition of TLNs in GR 
and revisit the procedure presented in Ref. \cite{Barura:2024uog}, 
where the TLNs of static, spherically symmetric BHs 
were computed using the Green's function technique.
\subsection{Definition of relativistic tidal Love numbers in GR}

For a staitc, spherically symmetric metric of the form as Eq.~(\ref{metricref}), we can fix the BH horizon radius to $r_h = 1$. To achieve this, the original horizon radius $r_h$  can be reparameterized as
\begin{equation}
	r_h(\eta)=2M+\sum_{n=1}^{N}\eta^{n}c_n,
\end{equation}
where $\eta$ denotes the parameter characterizing the deviation from the Schwarzschild BH and is expected to be very small. 
For instance, $\eta = q$ in the Bardeen BH case. Substituting this expansion into the metric function $f(\eta)$ and expanding it in powers of $\eta$,
\begin{equation}
	\sum_{k=0}^{N} \eta^{k}\, \frac{f^{(k)}(0)}{k!} = 0 ,
\end{equation}
we can solve for the coefficients $c_{1},c_{2},c_{3},\dots$ order by order by comparing the coefficients of $\eta$ at each expansion order. When the horizon is fixed to $r_h = 1$, the mass $M$ cannot be an independent quantity and can be determined by $f(r_h)=0$. In some modified metrics, $M$ cannot be obtained in closed form, 
so we treat $M$ itself as a power series in the deformation parameter $\eta$,
\begin{equation}\label{massexpansion}
	M(\eta)=\sum_{n=0}^{N}M_n\,\eta^{n}.
\end{equation}
Plugging this into  the condition $r_h(\eta)=1$ and expanding this equation in powers of $\eta$
\begin{equation}
	\sum_{k=0}^{N} \eta^{k}\, \frac{r_h^{(k)}(0)}{k!} = 1,
\end{equation}
we can sequentially solve for the mass expansion coefficients $M_0, M_1, M_2, \dots$.
In the subsequent calculations, we only retain terms up to the third order, i.e., $N=3$.
This procedure ensures that the horizon is always normalized at $r_h = 1$ in different modified gravity backgrounds, 
thereby providing a consistent framework for subsequent perturbation equations and the computation of TLNs.

Assuming that the external tidal field is weak, the linear perturbation can be used to study the tidal response of a self-gravitating object. Due to the spherical symmetry of the Schwarzschild spacetime, massless perturbations of integer spin $s$ admit complete separation of variables, leaving a single radial master function to be determined. In particular, the perturbations of scalar fields ($s=0$), electromagnetic fields ($s=1$), and the odd-parity (axial) gravitational perturbations ($s=2$) can all be written in a Schr\"odinger-like Regge-Wheeler (RW)-type equation, each characterized by a different effective potential.  In the gravitational case, the axial perturbations are governed by the original RW equation \cite{Regge:1957td}. In contrast, the even-parity (polar) gravitational perturbations of a Schwarzschild BH satisfy a distinct Schr\"odinger-like master equation, known as the Zerilli equation \cite{Zerilli:1970se}.
In the static limit, the RW equation reads
\begin{equation}\label{RWequation}
	 f_S(r) \frac{d}{dr} \left( f_S(r) \frac{d}{dr} {\Psi}(r)\right) - f_S(r) V_{\rm RW}(r)  {\Psi}(r) = 0, 
\end{equation}
where the effective potential is
\begin{equation}
	V_{\rm RW}(r) = \frac{l(l+1)}{r^2} + \frac{1-s^2}{r^3},
\end{equation}
the metric function for the Schwarzschild BH is $f_S(r) = 1 - 1/r$ and $s$ denotes the spin of the perturbing fields. It is known that the general solution to Eq.~(\ref{RWequation}) can be expressed in terms of hypergeometric functions, 
and it is straightforward to find the solution that is regular at the horizon. 
The asymptotic behavior of the solution at large $r$ is generally given by \cite{Hui:2020xxx}
\begin{equation}\label{RWsolution}
	\Psi(r) \sim r^{l+1} \big[ 1 + \mathcal{O}(r^{-1}) \big] 
	+ K_l^s \, r^{-l} \big[ 1 + \mathcal{O}(r^{-1}) \big],
\end{equation}
where $K_l^s$ is a constant that encodes the tidal response of the BH, which directly corresponds to the TLNs. This solution can be interpreted as consisting of two parts in the asymptotic region $r \to \infty$: the growing mode, which represents the external tidal field, and the decaying mode, which describes the induced response. It has been rigorously proven that the  TLNs of the Schwarzschild BH is vanishing~\cite{Fang:2005qq,Binnington:2009bb,Damour:2009vw,Kol:2011vg, Gurlebeck:2015xpa,Hui:2020xxx}. This finding has since been extended to Reissner-Nordström~\cite{Cardoso:2017cfl,Pereniguez:2021xcj,Rai:2024lho}, Kerr~\cite{Landry:2015zfa,Pani:2015nua,Poisson:2014gka,LeTiec:2020bos,LeTiec:2020spy,Chia:2020yla,Charalambous:2021kcz,Charalambous:2021mea,Ivanov:2022qqt} and Kerr-Newman~\cite{Ma:2024few}BHs.

The TLNs defined in Eq.~\eqref{RWsolution} are derived from the asymptotic expansion of linear perturbations at large distances, where the expansion consists solely of power-law terms in $r$. These values are constant and hence independent of the measurement scale. However, in the presence of matter fields~\cite{Cardoso:2019upw}, or within alternative backgrounds or modified gravity theories~\cite{DeLuca:2022tkm,Cardoso:2017cfl,Kol:2011vg}, the perturbation fields at large distances do not necessarily exhibit a simple power-law series. Instead, the asymptotic behavior may involve logarithmic terms,
\begin{equation}
	\begin{split}
		\Psi(r) \sim \;& r^{\,l+1}\bigl[1 + \mathcal{O}(r^{-1})\bigr] \\
		&+ K_{l}^{s} \ r^{-l} \,\bigl[\ln r + \mathcal{O}(1)\bigr]\bigl[1 + \mathcal{O}(r^{-1})\bigr].
	\end{split}
\end{equation}
The coefficient of the logarithmic term can be interpreted as a 
beta function in the context of a classical renormalization flow~\cite{Hui:2020xxx,Cardoso:2017cfl,Kol:2011vg}. 

\subsection{The Green’s function technique}

In the following we give a brief review of the Green’s function technique adopted in Ref.\cite{Barura:2024uog} to extract 
the TLNs of BHs in modified gravities. Assume that the static limit of the perturbation equations shares the form of the RW equation (\ref{RWequation}) but with distinct functions for the effective potential $V(r)$ and the metric $f(r)$. Performing the following transformation 
\begin{align}
	\Psi(r) =  \frac{\Phi(r)}{\sqrt{Z(r)}}, \quad
	Z(r) = \frac{f(r)}{f_{S}(r)},
\end{align}
 the perturbation equation becomes 
\begin{equation}\label{PerturbationEqModi}
	f_S(r)\,\frac{d}{d{r}}\!\Big(f_S(r)\,\frac{d}{d{r}}\,\Phi(r)\Big) 
	- f_S(r)\,U(r)\,\Phi(r) = 0.
\end{equation}
We can find that all the information of the modified spacetime  is encoded only in an effective potential
\begin{equation}\label{Ur}
	U(r) = \frac{V}{Z} 
	- \frac{1}{4 Z^2} \Big[ f_S \Big(\frac{d Z}{dr}\Big)^2 
	- 2 Z \frac{d}{dr} \Big(f_S\,\frac{d Z}{dr}\Big) \Big].
\end{equation}
In this work, we mainly consider the scalar, vector, and axial gravitational perturbations of RBHs. 
For the scalar and vector fields, the effective potentials are uniformly given by \cite{Konoplya:2024lch}
\begin{equation}\label{Vrscalarvector}
	V(r) = \frac{l(l+1)}{r^2} + \frac{1 - s^2}{r}\, f'(r).
\end{equation}
The case of axial gravitational perturbations is more complicated than that of scalar and vector fields. This is primarily because the former involves perturbing the underlying equations of motion.  However, since the metric (\ref{metricref}) is introduced as a phenomenological model---rather than being derived from a known action or as an exact solution to specific field equations---the underlying fundamental theory remains unknown. To address this issue, we adopt the viewpoint that quantum or regularization corrections  are effectively modeled as arising from the stress-energy tensor of an anisotropic fluid within the framework of classical Einstein gravity \cite{Toshmatov:2017kmw,Ashtekar:2018cay,Chen:2019iuo}. In this framework, the axial  perturbations are particularly simple, as they decouple from the matter perturbations. As a result, the perturbation equation is of the standard form given by \eqref{RWequation}, and the effective potential  is expressed as  ~\cite{Shi:2025gst} \footnote{It is important to note that the effective potential derived in~\cite{Shi:2025gst} applies to an anisotropic fluid with an arbitrary nonzero radial pressure. In contrast, the effective potential in \cite{Cardoso:2022whc} was obtained specifically for the case of vanishing background radial pressure. In the limit of $p_r^{(0)}=0$, the $rr$ component of the Einstein field equations yields the following relation for the metric gradient: $a'(r)=\frac{a(r)}{r}\frac{1-b(r)}{b(r)}$. Using this relation, Eq. (\ref{Vrofg}) consistently  reproduces the results presented in \cite{Cardoso:2022whc} when the limit $p_r^{(0)} = 0$ is taken.}

\begin{equation}\label{Vrofg}
	V(r) = \frac{l(l+1)}{r^2} - \frac{f'(r)}{r} + f''(r).
\end{equation}
 Next, we consider the following expansions for the effective potential and the master function:
\begin{equation}
	U(r) = \sum_{k \ge 0} \eta^k \, U^{(k)}(r), 
	\quad \Phi(r) = \sum_{k \ge 0} \eta^k \, \Phi^{(k)}(r),
\end{equation}
where as before the parameter $\eta$ represents the deviation from the Schwarzschild BH. Obviously, the zeroth-order potential is $U^{(0)}(r) = V_{\rm RW}(r)$.

Our strategy is to solve Eq. (\ref{PerturbationEqModi}) order by order in $\eta$ by the Green’s function method. At the leading order of $\eta$, we recover the standard RW equation.
\begin{equation}\label{RWeqU0}
	\Big[ f_S(r) \frac{d}{dr} \Big( f_S(r) \frac{d}{dr} \Big) - f_S(r) U^{(0)}(r) \Big] \, \Phi^{(0)}(r) = 0,
\end{equation}
This equation has three regular singular points and can be easily transformed into the standard hypergeometric equation. The two linearly independent solutions to this equation are written in terms of the hypergeometric function
as 
\begin{align}
	\Phi^{(0)}_{+}(r) &= r^{\,l+1}\,{}_2F_1\left(-l-s, -l+s; -2l; \frac{1}{r}\right), \label{eq:Phi_plus} \\
	\Phi^{(0)}_{-}(r) &= r^{-l}\,{}_2F_1\left(l+1-s, l+1+s; 2l+2; \frac{1}{r}\right).\label{eq:Phi_minus}
\end{align}
As analyzed in Refs. \cite{Barura:2024uog, Motaharfar:2025ihv}, for $l\in \mathbb{C}$, both the solutions are divergent in the form $\propto \ln(r-1)$ at the horizon. However, a unique linear combination of the two solutions exactly cancels the logarithmic terms  at the horizon. As a result, the horizon-regular solution can be written as
\begin{equation}\label{eq:Phi_hor_reg}
	\Phi^{(0)}(r)=\Phi^{(0)}_{\text{hor-reg}}(r) = \Phi^{(0)}_{+}(r) + K_l^{s(0)}\, \Phi^{(0)}_{-}(r),
\end{equation}
Here, $K_l^{s(0)}$ is the constant encoding the tidal response at zero order, whose explicit expression is
\begin{equation}
	K_l^{s(0)} = - \frac{\Gamma(-2l)\, \Gamma(l+1+s)\, \Gamma(l+1-s)}{\Gamma(-l-s)\, \Gamma(-l+s)\, \Gamma(2l+2)}.
\end{equation}
This value is zero for  a physical multipole $l \in \mathbb{Z}$, which corresponds to  the fact that the TLNs are vanishing for four-dimensional BHs in GR.

Next, we consider the higher-order expansion of the perturbation equation \eqref{PerturbationEqModi}.  At $k$-th order, $\mathcal{O}(\eta^k)$,  $k\geq1$, the equation takes the form
\begin{equation}\label{InhomogEq}
	\begin{split}
		f_S(r) \frac{d}{dr} \Bigl( f_S(r) \frac{d}{dr} \Phi^{(k)}(r) \Bigr) 
		- f_S(r) U^{(0)}(r) \, \Phi^{(k)}(r) \\
		= f_S(r) \, \mS^{(k)}(r),
	\end{split}
\end{equation}
where the source term $\mS^{(k)}(r)$ is given by
\begin{equation}\label{SKr}
	\mS^{(k)}(r) = \sum_{i=0}^{k-1} U^{(k-i)}(r) \, \Phi^{(i)}(r).
\end{equation}
By observing that the derivative operator of the above inhomogeneous equation is the same as that of the homogeneous equation \eqref{RWeqU0}, the solution to the inhomogeneous equation can be obtained using the Green's function method. The Green's function $G(r, r')$ is defined by
\begin{equation}
	\frac{d}{dr} \Bigl( f_S(r) \frac{dG(r,r')}{dr} \Bigr) - U^{(0)}(r)\, G(r,r') = \delta(r-r'),
\end{equation}
with $\delta(r)$ being the  delta function. Here, appropriate boundary conditions must be imposed. At the horizon $r = 1$, the field $\Phi(r)$ is required to be regular. As $r \to \infty$, we demand that $\Phi(r) = \mathcal{O}(r^l)$,
which is justified by the fact that the full-order solution $\Phi(r)$ can always be renormalized such that the coefficient of the growing mode $r^{\,l+1}$ is set to unity. Then the Green's function can be constructed with the two linearly independent solutions \eqref{eq:Phi_plus} and \eqref{eq:Phi_minus} of the corresponding homogeneous equation \eqref{RWeqU0} 
\begin{align}
	G(r, r') &= -\frac{1}{2l+1} \Bigl[ 
	\Phi^{(0)}_-(r)\, \Phi_{\rm hor\text{-}reg}(r') \, \Theta(r-r') \notag\\
	&\qquad + \Phi_{\rm hor\text{-}reg}(r)\, \Phi^{(0)}_-(r') \, \Theta(r'-r)
	\Bigr],
\end{align}
where $\Theta(r)$ denotes the Heaviside step function and the factor $2l+1$ arises from the Wronskian of the homogeneous solutions \cite{Barura:2024uog}. 

Using this Green's function, the $k$th-order solution of the inhomogeneous equation \eqref{InhomogEq} can be written as
\begin{align}\label{Inhomogsolu}
	\Phi^{(k)}(r) &= \int_1^{\infty} G(r,r') \, \mS^{(k)}(r') \, dr' \notag\\
	&= -\frac{1}{2l+1} \Biggl[
	\Phi^{(0)}_-(r) \int_1^r \Phi_{\rm hor\text{-}reg}(r') \mS^{(k)}(r') \, dr' \notag\\
	&\qquad + \Phi_{\rm hor\text{-}reg}(r) \int_r^{\infty} \Phi^{(0)}_-(r') \mS^{(k)}(r') \, dr'
	\Biggr].
\end{align}
With this solution, the TLNs can be determined by 
the coefficient of the decaying mode $r^{-l}$ at large $r$. 
As pointed out in Ref. \cite{Barura:2024uog}, for $l \in \mathbb{Z}$,  only the constant part of the integral contributes 
to this term in the $r \to \infty$ limit.  Accordingly, the TLNs are identified from the constant term 
in the following expression:
\begin{equation}\label{ISkregular}
	I\bigl[\mS^{(k)}(r)\bigr] 
	= -\frac{1}{2l+1} \int_{1}^{r} 
	\Phi^{(0)}_{+}(r') \, \mS^{(k)}(r') \, dr' .
\end{equation}
We emphasize that when logarithmic behavior appears in the asymptotic expansion, the logarithmic terms are also regarded as part of the TLNs.

\section{TIDAL LOVE NUMBERS FOR REGULAR BLACK HOLES}\label{sectionIV}
In this section, we investigate the linear static response of the three types of RBHs introduced in Sec.~\ref{backgroung} under the influence of external tidal fields (scalar, vector, and axial gravitational types). By employing the Green’s function method introduced in Sec.~\ref{sectionIII}, we can extract the TLNs.

\subsection{Bardeen black hole}
Here, we extract the response of the Bardeen BH under scalar, vector, and axial gravitational perturbations. Since we set the horizon radius to $r_h = 1$, the mass parameter can be expanded as  
\begin{equation}
	M(q) = \tfrac{1}{2} + \tfrac{3}{4}q^2 + \tfrac{3}{16}q^4 - \tfrac{1}{32}q^6 + \mathcal{O}(q^8).
\end{equation}
Correspondingly, the metric function $f(r)$ admits a series expansion in powers of $q$ given by  
\begin{equation}\label{frBardeen}
	\begin{aligned}
		f(r) &= 1 - \frac{1}{r}- \frac{3 q^2 (r^2 - 1)}{2 r^3} - 2 q^4 \Biggl( \frac{3}{16 r} - \frac{9}{8 r^3} + \frac{15}{16 r^5} \Biggr) \\
		&\quad - 2 q^6 \Biggl( \frac{1}{32 r} + \frac{9}{32 r^3} - \frac{45}{32 r^5} + \frac{35}{32 r^7} \Biggr) 
		+ \mathcal{O}(q^8).
	\end{aligned}
\end{equation}
Plugging this into Eq.\eqref{Vrscalarvector} and then Eq. \eqref{Ur}, and performing the expansion in powers of $q$, the power-series expansion of the effective potential for the scalar ($s=0$) and vector ($s=1$) perturbations can be collectively expressed as
	\begin{align}
		U_{1}(r) &= 
		\frac{-3 + 6\,l(l+1) - 6\,s^{2}}{4\,r^{4}} 
		+ \frac{6\,l(l+1) - 6\,s^{2}}{4\,r^{3}}\notag\\
		&+ \frac{12 + 12\,s^{2}}{4\,r^{5}}, \label{eq:U1}\\[4pt]
		U_{2}(r) &= 
		\frac{3\,l(l+1) - 3\,s^{2}}{8\,r^{3}} 
		+ \frac{-6 + 21\,l(l+1) - 21\,s^{2}}{8\,r^{4}}\notag\\
        &+ \frac{57 + 42\,l(l+1) + 30\,s^{2}}{16\,r^{5}} 
		+ \frac{99 + 6\,l(l+1) + 66\,s^{2}}{16\,r^{6}}\notag\\
		&\quad 
		- \frac{33 + 12\,s^{2}}{4\,r^{7}}, \label{eq:U2}\\[4pt]
		U_{3}(r) &= 
		-\frac{l(l+1) - s^{2}}{16\,r^{3}}
		+ \frac{-4 + 17\,l(l+1) - 17\,s^{2}}{16\,r^{4}}\notag\\
		&+ \frac{-43 + 160\,l(l+1) - 124\,s^{2}}{32\,r^{5}} \notag\\
		&+ \frac{219 + 160\,l(l+1) + 92\,s^{2}}{32\,r^{6}}\notag\\
		& 
		+ \frac{-41 + 34\,l(l+1) + 74\,s^{2}}{32\,r^{7}} \notag\\
		&- \frac{391 + 2\,l(l+1) + 106\,s^{2}}{32\,r^{8}}\notag\\
		&
		+ \frac{27 + 6\,s^{2}}{2\,r^{9}}. \label{eq:U3}
	\end{align}
For the axial gravitational perturbation ($s=2$), plugging Eq. \eqref{frBardeen} into Eq. \eqref{Vrofg} and then Eq. \eqref{Ur}, and performing the expansion in powers of $q$, one obtains the effective-potential expansion coefficients
\begin{align} U_{1}(r) &= \frac{-9 + 3\,l(l + 1)}{2 r^{3}} + \frac{-21 + 6\,l(l + 1)}{4 r^{4}} + \frac{21}{r^{5}},\\[4pt] 
	U_{2}(r) &= \frac{-9 + 3\,l(l + 1)}{8 r^{3}} + \frac{-69 + 21\,l(l + 1)}{8 r^{4}}\notag \\ &\quad + \frac{363 + 42\,l(l + 1)}{16 r^{5}} + \frac{513 + 6\,l(l + 1)}{16 r^{6}}\notag\\ &\quad - \frac{165}{4 r^{7}},
\end{align}  
\begin{align}
	U_{3}(r) &= -\,\frac{-3 + l + l^{2}}{16 r^{3}} + \frac{-55 + 17\,l + 17\,l^{2}}{16 r^{4}} \notag\\ &\quad + \frac{-307 + 160\,l(1 + l)}{32 r^{5}} + \frac{1251 + 160\,l(1 + l)}{32 r^{6}} \nonumber\\ &\quad + \frac{-215 + 34\,l(1 + l)}{32 r^{7}} - \frac{1753 + 2\,l(1 + l)}{32 r^{8}}\notag\\ &\quad + \frac{123}{2 r^{9}}.
\end{align}
\subsubsection{Scalar field response}
In this section, we compute the TLN of Bardeen BH under scalar perturbation.  The lowest multipole number starts from $l=0$. 
For \(l = 0\), the zeroth-order \(\mathcal{O}(q^{0})\) growing and decaying modes are given by Eqs.~(\ref{eq:Phi_plus}) and (\ref{eq:Phi_minus}), and take the following form:
\begin{align}
\Phi^{(0)}_{+}(r)
&= \Phi^{(0)}(r)
= \Phi_{\mathrm{hor\text{-}reg}}(r)
= r , \\[2mm]
\Phi^{(0)}_{-}(r)
&= -\, r \ln\!\left(1-\frac{1}{r}\right) .
\end{align}

We should point out that the first equality in the first equation follows from Eq.~(\ref{eq:Phi_hor_reg}), 
which relies on the fact that $K_{l}^{s}$ vanishes for integer multipole number $l$ 
in the case of the classical Schwarzschild BH. 
For $s = 0$ and $l = 0$, the source terms at first order 
$\mathcal{O}(q^{2})$ can be obtained from Eq.~(\ref{SKr}), yielding
\begin{align}
S^{(1)}(r) &= U^{(1)}(r)\, \Phi^{(0)}(r)=\frac{3}{r^{4}} - \frac{3}{4 r^{3}}.
\end{align}
Furthermore, the first-order solution and the second-order source term at $\mathcal{O}(q^{4})$ can be written as follows:
\begin{align}
\Phi^{(1)}(r)
&= \int_{1}^{\infty} G(r,r')\, S^{(1)}(r')\, dr'
= -\frac{3}{4} - \frac{3}{4r},
\\[3mm]
S^{(2)}(r)
&= U^{(2)}(r)\,\Phi^{(0)}(r) + U^{(1)}(r)\,\Phi^{(1)}(r)
\notag \\
          &= -\frac{21}{2 r^{6}} + \frac{9}{2 r^{5}} + \frac{33}{8 r^{4}} - \frac{3}{4 r^{3}}.
\end{align}
Moreover, the second-order solution and the third-order source term at $\mathcal{O}(q^{6})$, following Eqs.~(\ref{Inhomogsolu}) and (\ref{SKr}), can be computed as follows:
\begin{align}
\Phi^{(2)}(r)
&= \int_{1}^{\infty} G(r,r')\, S^{(2)}(r')\, dr'\notag \\
        &= -\frac{3}{16} + \frac{21}{32 r^{3}} + \frac{3}{8 r^{2}} - \frac{15}{32 r},
\\[3mm]
S^{(3)}(r)
&= U^{(3)}(r)\, \Phi^{(0)}(r)
   + U^{(2)}(r)\, \Phi^{(1)}(r)\notag \\
   &\quad + U^{(1)}(r)\, \Phi^{(2)}(r)\notag \\
        &= \frac{693}{32 r^{8}} - \frac{1285}{128 r^{7}}
            - \frac{329}{32 r^{6}} + \frac{579}{128 r^{5}}
            - \frac{41}{64 r^{4}} - \frac{1}{4 r^{3}}.
\end{align}

For $l=0$, the TLN is obtained from Eq.~(\ref{ISkregular}), which expanded in $q^2$ gives
\begin{align}
		I[\mS^{(1)}](r) &= -\frac{3}{4} + \frac{3}{2 r^2} - \frac{3}{4 r}, \\
		I[\mS^{(2)}](r)  &= -\frac{3}{16} - \frac{21}{8 r^4} + \frac{3}{2 r^3} 
		+ \frac{33}{16 r^2} - \frac{3}{4 r}, \\
		I[\mS^{(3)}](r)  &=\frac{1}{32} + \frac{231}{64 r^6} - \frac{257}{128 r^5} 
		- \frac{329}{128 r^4} + \frac{193}{128 r^3}
		\notag \\
        &\quad- \frac{41}{128 r^2}- \frac{1}{4 r}.
\end{align}
We find that, in the case of the scalar perturbation with $l=0$, 
there exist nonvanishing TLNs and no logarithmic terms appear. Moreover, we find that the first- and second-order terms provide negative contributions to the TLNs, 
while the third-order term gives a positive contribution. Up to $\mathcal{O}(q^{8})$, the TLN is given by
\begin{equation}
	K_{l=0}^{0} = 
	- \frac{3}{4} q^{2} 
	- \frac{3}{16} q^{4} 
	+ \frac{1}{32} q^{6} 
	+ \mathcal{O}(q^{8}) .
\end{equation}

For $l=1$, the zeroth-order solutions~(\ref{eq:Phi_plus}) and~(\ref{eq:Phi_minus}) take the following form:
\begin{align}
\Phi^{(0)}_{+}(r)
&= \Phi^{(0)}(r)
= \Phi_{\mathrm{hor\text{-}reg}}(r)
= -\frac{r}{2} + r^2 , \\[2mm]
\Phi^{(0)}_{-}(r)
&= -12\, r + 6\, r \, \ln\!\left(1 - \frac{1}{r}\right)
   - 12\, r^2 \, \ln\!\left(1 - \frac{1}{r}\right) .
\end{align}
The corresponding integrals Eq.~(\ref{ISkregular}) expanded in $q^2$ are given by

	\begin{align}
		I[\mS^{(1)}](r) &= \frac{15}{16} + \frac{1}{8r^2} - \frac{13}{16r} + \frac{r}{4} - \frac{r^2}{2} - \frac{\ln r}{2}, \\[4pt]
		I[\mS^{(2)}](r)  &= -\frac{255}{64} - \frac{15 \pi^2}{16} - \frac{7}{32 r^4} + \frac{3}{2 r^3} 
		- \frac{305}{64 r^2}\notag\\
        &- \frac{\pi^2}{8 r^2} + \frac{331}{32 r} + \frac{13 \pi^2}{16 r} - \frac{11 r}{4} - \frac{\pi^2 r}{4} - \frac{r^2}{8}\nonumber\\
        &+ \frac{\pi^2 r^2}{2}- \frac{3 \ln r}{4} 
		+ \pi^2 \ln r - \frac{15 \ln r}{8 r^2} + \frac{63 \ln r}{8 r}\nonumber\\
        &- 3 r \ln r - \frac{45 \ln^2 r}{16} - \frac{3 \ln^2 r}{8 r^2} + \frac{39 \ln^2 r}{16 r} \nonumber\\
		& - \frac{3}{4} r \ln^2 r + \frac{3}{2} r^2 \ln^2 r + \frac{\ln^3 r}{2} - \frac{45}{8} \operatorname{Li}_2(1 - r)\nonumber\\
        &- \frac{3 \operatorname{Li}_2(1 - r)}{4 r^2} 
		+ \frac{39 \operatorname{Li}_2(1 - r)}{8 r}
		 - \frac{3}{2} r \operatorname{Li}_2(1 - r)\nonumber\\
         &+ 3 r^2 \operatorname{Li}_2(1 - r)
         + 3 \ln r \operatorname{Li}_2(r) - 6 \operatorname{Li}_3(r)\nonumber\\
         &+ 6 \zeta(3) .
	\end{align}
We can see that for $l=1$, the TLNs is nonzero and logarithmic terms appear.  In particular, the $I[\mS^{(1)}]$ part is simple enough to extract TLN explicitly: The corresponding coefficient reads
$K_{l=1}^{0\,(1)} = {15}/{16} - {1}/{2} \ln r$. 
However, the $I[\mS^{(2)}]$ part is more complicated. This complexity arises from the multiple integrations of logarithmic terms during the computation. Such integrations unavoidably produce the polylogarithm function. Nevertheless, we can extract the constant and $\ln r$ terms from $I[\mathcal{S}^{(2)}](r)$ at large $r$. The TLN at the second order of $q^2$ is then obtained as $K_{l=1}^{0\,(2)} = -{303}/{64} + 6\,\zeta(3) - {3}/{4}\ln r$, which is positive and has a negative logarithmic
running at the second order.
Note that we compute the TLN only up to second order, since the integration \eqref{Inhomogsolu} cannot be performed analytically further, and consequently, we cannot construct the third-order source term. Up to $\mathcal{O}(q^{6})$, the TLN is given by
\begin{align}
K_{l=1}^{0} &=
\left(\frac{15}{16} - \frac{1}{2}\ln r \right) q^{2}\notag\\
        &+ \left( -\frac{303}{64} + 6\,\zeta(3) - \frac{3}{4}\ln r \right) q^{4}+ \mathcal{O}(q^{6}) .
\end{align}

\subsubsection{Vector field response}
Now we compute the TLNs of the Bardeen BH under vector perturbation. Similar to the steps present in the previous subsection, for $s=1$ and $l=1$, the TLNs can be obtained from the following expressions:
	\begin{align}
		I[\mS^{(1)}](r) &= \frac{1}{2} - \frac{r}{4} - \frac{r^2}{4} - 2 \ln r, \\[4pt]
		I[\mS^{(2)}](r) &= -\frac{411}{16} - \frac{\pi^2}{2} + \frac{3}{8 r^2} + \frac{217}{8 r} - \frac{7 r}{4}
		+ \frac{\pi^2 r}{4} \notag\\
        &- \frac{r^2}{16} + \frac{\pi^2 r^2}{4} - \frac{21 \ln r}{4}
		+ 4 \pi^2 \ln r + \frac{3 \ln r}{r^2} \notag\\
        &+ \frac{51 \ln r}{4 r}  - \frac{3}{2} r \ln r - \frac{3 \ln^2 r}{2}  
		+ \frac{3}{4} r \ln^2 r \notag\\
        &+ \frac{3}{4} r^2 \ln^2 r + 2 \ln^2 r 
		- 3 \mathrm{Li}_2(1 - r) + \frac{3}{2} r\, \mathrm{Li}_2(1 - r) \notag\\
		& + \frac{3}{2} r^2\, \mathrm{Li}_2(1 - r)
		+ 12 \ln r\, \mathrm{Li}_2(r) - 24 \mathrm{Li}_3(r)\notag\\
        &+ 24 \zeta(3).
	\end{align}
From $I[\mS^{(1)}](r)$, we obtain the first-order vector TLN $K_{1}^{1\,(1)} = {1}/{2} - 2 \ln r $, which is positive and has a negative logarithmic running. 
From the asymptotic behavior of $I[\mS^{(1)}](r)$ at large $r$,  the second-order vector TLN is obtained as 
$K_{1}^{1\,(2)} = -{381}/{16} + 24\,\zeta(3) - 3 \ln r$, which is negative plus a positive logarithmic running. Up to $\mathcal{O}(q^{6})$, the vector TLN for $l=1$ is given by
\begin{align}
K_{l=1}^{1} &=
\left( \frac{1}{2} - 2 \ln r \right) q^{2}\notag\\
        &+ \left( -\frac{381}{16} + 24\,\zeta(3) - 3 \ln r \right) q^{4}+ \mathcal{O}(q^{6}) .
\end{align}

Finally, for $l=2$, the  integration \eqref{ISkregular} for vector perturbation at each order of $q$ becomes 
	\begin{align}
		I[\mS^{(1)}](r)  &= -\frac{153}{160} + \frac{333 r}{320} - \frac{3 r^2}{320} 
		+ \frac{3 r^3}{10} - \frac{3 r^4}{8} - \frac{27 \ln r}{40}, \label{IS1}\\[4pt]
		I[\mS^{(2)}](r)  &= -\frac{9597}{256} + \frac{1377 \pi^2}{160} + \frac{1161}{640 r^2} 
		+ \frac{25713}{640 r} + \frac{2829 r}{640} \notag\\
        &- \frac{2997 \pi^2 r}{320}  
		+ \frac{2847 r^2}{256} + \frac{27 \pi^2 r^2}{320} - \frac{399 r^3}{20}\notag\\
		&- \frac{3 r^4}{32} - \frac{27 \pi^2 r^3}{10}  + \frac{27 \pi^2 r^4}{8} + \frac{18117 \ln r}{320}\notag\\ 
		&+ \frac{243 \pi^2 \ln r}{20} + \frac{81 \ln r}{80 r^2} + \frac{3753 \ln r}{320 r} 
		+ \frac{27 r \ln r}{32}\notag\\
		&+ \frac{243 r^2 \ln r}{40}   - \frac{81 r^3 \ln r}{4} + \frac{4131 \ln^2 r}{160} \notag\\
        &
		- \frac{8991 r \ln^2 r}{320} + \frac{81 r^2 \ln^2 r2}{320} 
		- \frac{81 r^3 \ln^2 r}{10}\notag\\
		&+ \frac{81 r^4 \ln^2 r}{8}  + \frac{243 \ln^3 r}{40}
		+ \frac{4131}{80} \mathrm{Li}_2(1 - r) \notag\\
		&+ \frac{729 \zeta(3)}{10} - \frac{8991 r}{160} \mathrm{Li}_2(1 - r)+ \frac{81 r^2}{160} \mathrm{Li}_2(1 - r) \notag\\
        &- \frac{81 r^3}{5} \mathrm{Li}_2(1 - r) 
		+ \frac{81 r^4}{4} \mathrm{Li}_2(1 - r) 
		\notag\\&+ \frac{729}{20} \ln r\, \mathrm{Li}_2(r)
		- \frac{729}{10} \mathrm{Li}_3(r). \label{IS2}
	\end{align}
The first-order correction of the vector TLN directly reads
$K_{2}^{1\,(1)} = -{153}/{160} - {27}/{40}\ln r$. The second-order correction of the vector TLN is obtained from the asymptotic behavior of $I[\mS^{(1)}](r)$ at large $r$, which is given by
$K_{2}^{1\,(2)} = -{24087}/{256} + {729}/{10}\,\zeta(3) + {27}/{80} \ln r $. Up to $\mathcal{O}(q^{6})$, the vector TLN  for $l=2$ turns out to be
\begin{align}
K_{l=2}^{1} &=
\left( -\frac{153}{160} - \frac{27}{40} \ln r \right) q^{2}\notag\\
        &+ \left( -\frac{24087}{256} + \frac{729}{10}\,\zeta(3) + \frac{27}{80} \ln r \right) q^{4}+ \mathcal{O}(q^{6}) .
\end{align}

\subsubsection{Axial gravitational field response}

Here, we compute the TLNs of the Bardeen BH 
under axial gravitational perturbations for \(l=2\) and \(l=3\). 
We carry the calculations up to second order in $q^4$. For \(l=2\), the expansion of the integration \eqref{ISkregular} in $q^4$ reads
\begin{align}
	I[\mS^{(1)}](r) &= \frac{103}{40} - \frac{21\,r^2}{10} - \frac{r^3}{4} - \frac{9\,r^4}{40}, \\[4pt]
	I[\mS^{(2)}](r) &= \frac{3541}{128} - \frac{2163}{320\,r^2} - \frac{13081}{640\,r} + \frac{837\,r}{320}
	- \frac{231\,r^2}{80}\notag \\
	&\quad - \frac{11\,r^3}{80} - \frac{9\,r^4}{160} + \frac{6193}{320}\,\ln r.
\end{align}
From these results we see that the first-order TLN is nonvanishing and contains no logarithmic running, 
whereas logarithmic running appears at second order through the \(\ln r\) term in \(I[\mS^{(2)}](r) \). In this case, all the correction terms contribute 
positively to the TLN.
The TLN up to $\mathcal{O}(q^6)$ can be expressed as
\begin{equation}
	K_{l=2}^{2} =
	\frac{103}{40}\, q^{2}
	+ \left( \frac{3541}{128} + \frac{6193}{320}\, \ln r \right) q^{4}
	+ \mathcal{O}(q^{6}) .
\end{equation}
For the case of $l = 3$, we obtain the following results:
\begin{align}
	I[\mS^{(1)}](r)  &= 
	\frac{1667}{10080}
	- \frac{25 r^{2}}{24}
	+ \frac{1255 r^{3}}{1008}
	- \frac{73 r^{4}}{224}
	+ \frac{39 r^{5}}{140}\notag \\
	&\quad
	- \frac{9 r^{6}}{28},\\[2mm]
	I[\mS^{(2)}](r)  &= 
	\frac{1934113}{161280}
	- \frac{1667}{3840\, r^{2}}
	- \frac{1918717}{806400\, r}
	- \frac{600521\, r}{44800}\notag \\
	&\quad
	+ \frac{2393\, r^{2}}{2688}
	+ \frac{70667\, r^{3}}{20160}
	- \frac{2081\, r^{4}}{4480}
	+ \frac{3\, r^{5}}{8}\notag \\
	&\quad
	- \frac{9\, r^{6}}{112}
	+ \frac{18889\, \ln r}{403200}.
\end{align}
Similarly, the TLN for $l = 3$ up to $\mathcal{O}(q^6)$ turns out to be
\begin{equation}
	K_{l=3}^{2} =
	\frac{1667}{10080}\, q^{2}
	+ \left( \frac{1934113}{161280} + \frac{18889}{403200}\, \ln r \right) q^{4}
	+ \mathcal{O}(q^{6}) .
\end{equation}
The behavior of TLN for $l=3$ closely resembles that of the $l = 2$ case: 
The first-order term  contributes finite and nonzero TLNs without logarithmic running, 
while the second-order correction  introduces a logarithmic dependence in $r$. Both cases contribute positively to the TLNs and the running.

\subsection{Regular black holes with sub-Planckian curvature}
In this subsection, we compute the TLNs of the RBH with sub-Planckian curvature ~\cite{Ling:2021olm}. Without loss of generality, we choose the parameters $x = 1$ and $c = 3$.
Similar to the Bardeen BH case, we consider the response under scalar, vector, and axial gravitational perturbations. Setting the horizon radius $r_h = 1$, the mass can be expanded according to Eq.~\eqref{massexpansion} as
\begin{equation}
	M(\alpha_0) = \frac{1}{2} + \frac{\alpha_0}{4} + \frac{3 \, \alpha_0^2}{16} + \frac{\alpha_0^3}{6} + \mathcal{O}(\alpha_0^4) \,.
\end{equation}
Up to the same order in the regular parameter $\alpha_0$, the metric function can be expanded as 
\begin{align}
	f(r) &= 1 - \frac{1}{r} 
	+ \frac{(1 - r^3) \, \alpha_0}{2 \, r^4} 
	+ \frac{(-1 + 4 r^3 - 3 r^6) \, \alpha_0^2}{8 \, r^7} \notag\\
	&\quad
	+ \frac{(1 - 9 r^3 + 24 r^6 - 16 r^9) \, \alpha_0^3}{48 \, r^{10}} 
	+ \mathcal{O}(\alpha_0^4) \,.
\end{align}

The effective potentials \eqref{Vrscalarvector} for scalar ($s=0$) and vector ($s=1$) perturbations can be expanded in the same way as follows:
	\begin{align}
		U_1(r) &= 
		\frac{2\,l(l+1) - 2\,s^2}{4\,r^3} 
		+ \frac{-1 + 2\,l(l+1) - 2\,s^2}{4\,r^4}\notag\\
		&+ \frac{-2 + 2\,l(l+1) - 2\,s^2}{4\,r^5} 
		+ \frac{9 + 6\,s^2}{4\,r^6}, \label{U1}\\[2mm]
		U_2(r) &= 
		\frac{6\,l(l+1) - 6\,s^2}{16\,r^3} 
		+ \frac{-4 + 10\,l(l+1) - 10\,s^2}{16\,r^4} \notag\\
		&+ \frac{-11 + 14\,l(l+1) - 14\,s^2}{16\,r^5}  + \frac{27 + 10\,l(l+1) + 14\,s^2}{16\,r^6} \notag\\
		&+ \frac{14 + 6\,l(l+1) + 6\,s^2}{16\,r^7} 
		+ \frac{22 + 2\,l(l+1) + 10\,s^2}{16\,r^8} \notag\\
		&- \frac{9}{16\,r^9}, \label{U2}\\[2mm]
		U_3(r) &= 
		\frac{32\,l(l+1) - 32\,s^2}{96\,r^3} 
		+ \frac{-25 + 68\,l(l+1) - 68\,s^2}{96\,r^4} \notag\\
		&+ \frac{-83 + 116\,l(l+1) - 116\,s^2}{96\,r^5}
		+ \frac{51 + 64\,l(l+1) + 8s^2}{48\,r^6} \notag\\
		&+ \frac{86 + 116\,l(l+1) + 10\,s^2}{96\,r^7}
		+ \frac{217 + 80\,l(l+1) + 82s^2}{96\,r^8} \notag\\
		&+ \frac{30 + 19l(l+1) + 26s^2}{96\,r^9} 
		+ \frac{83 + 14\,l(l+1) + 40s^2}{96\,r^{10}}\notag\\
		&+ \frac{19 + 2\,l(l+1) + 16\,s^2}{96\,r^{11}}. \label{U3}
	\end{align}
The expansion of the effective potential \eqref{Vrscalarvector} for axial gravitational perturbations ($s=2$) is given as follows:
	\begin{align}
		U_{1}(r) &= 
		\frac{-3 + l(l+1)}{2\,r^{3}} 
		+ \frac{-7 + 2\,l(l+1)}{4\,r^{4}} \notag\\
		&+ \frac{-4 + l(l+1)}{2\,r^{5}} 
		+ \frac{51}{4\,r^{6}}, \label{U1}\\[1mm]
		U_{2}(r) &= 
		\frac{-9 + 3\,l(l+1)}{8\,r^{3}} 
		+ \frac{-17 + 5\,l(l+1)}{8\,r^{4}} \notag\\
		&+ \frac{-53 + 14\,l(l+1)}{16\,r^{5}} 
		+ \frac{165 + 10\,l(l+1)}{16\,r^{6}} \notag\\
		&+ \frac{40 + 3\,l(l+1)}{8\,r^{7}} 
		+ \frac{50 + l(l+1)}{8\,r^{8}} \notag\\
		&- \frac{45}{16\,r^{9}}, \label{U2}\\[1mm]
		U_{3}(r) &= 
		\frac{-3 + l(l+1)}{3\,r^{3}} 
		+ \frac{-229 + 68\,l(l+1)}{96\,r^{4}}\notag\\  
		&+ \frac{-431 + 116\,l(l+1)}{96\,r^{5}} 
		+ \frac{363 + 64\,l(l+1)}{48\,r^{6}} \notag\\
		& + \frac{155 + 29\,l(l+1)}{24\,r^{7}} 
		+ \frac{1111 + 80\,l(l+1)}{96\,r^{8}} \notag\\
		&+ \frac{126 + 19\,l(l+1)}{48\,r^{9}} 
		+ \frac{311 + 14\,l(l+1)}{96\,r^{10}}  \notag\\
		&\quad + \frac{31 + 2\,l(l+1)}{96\,r^{11}}. \label{U3}
	\end{align}

\subsubsection{Scalar field response}
First of all, we compute the TLNs for the RBH with sub-Planckian curvature under scalar field perturbations. 
For the case $l=0$, the zeroth-order (\(\mathcal{O}(\alpha_{0}^{0})\)) growing and decaying modes, corresponding to Eqs.~(\ref{eq:Phi_plus}) and (\ref{eq:Phi_minus}), are given explicitly by
\begin{align}
\Phi^{(0)}_{+}(r)
&= \Phi^{(0)}(r)
= \Phi_{\mathrm{hor\text{-}reg}}(r)
= r , \\[2mm]
\Phi^{(0)}_{-}(r)
&= -\, r \, \ln\!\left(1-\frac{1}{r}\right) .
\end{align}

In the case \(s = 0\) and \(l = 0\), the first-order source term at \(\mathcal{O}(\alpha_{0}^{1})\) is derived from Eq.~(\ref{SKr}) as
\begin{align}
S^{(1)}(r) &= U^{(1)}(r)\, \Phi^{(0)}(r)
= \frac{9}{4 r^5} - \frac{1}{2 r^4} - \frac{1}{4 r^3}.
\end{align}

Applying the Green’s function formalism, the first-order solution and the second-order source term (\(\mathcal{O}(\alpha_{0}^{2})\)) are expressed as
\begin{align}
\Phi^{(1)}(r)
&= \int_{1}^{\infty} G(r,r')\, S^{(1)}(r')\, dr'
= -\frac{1}{4} - \frac{1}{4 r^2} - \frac{1}{4 r}, \\[2mm]
S^{(2)}(r)
&= U^{(2)}(r)\,\Phi^{(0)}(r) + U^{(1)}(r)\,\Phi^{(1)}(r) \notag \\
&= -\frac{9}{8 r^8} + \frac{15}{16 r^7} + \frac{1}{2 r^6} + \frac{15}{8 r^5} - \frac{5}{8 r^4} - \frac{1}{4 r^3}.
\end{align}

Similarly, the second-order solution and the third-order source term at \(\mathcal{O}(\alpha_{0}^{3})\), following Eqs.~(\ref{Inhomogsolu}) and (\ref{SKr}), are given by
\begin{align}
\Phi^{(2)}(r)
&= \int_{1}^{\infty} G(r,r')\, S^{(2)}(r')\, dr' \notag \\
&= -\frac{3}{16} + \frac{1}{32 r^5} - \frac{1}{32 r^3} - \frac{1}{4 r^2} - \frac{7}{32 r}, \\[2mm]
S^{(3)}(r)
&= U^{(3)}(r)\, \Phi^{(0)}(r)
   + U^{(2)}(r)\, \Phi^{(1)}(r)
   + U^{(1)}(r)\, \Phi^{(2)}(r) \notag \\
&= \frac{27}{128 r^{11}} - \frac{1}{48 r^{10}} + \frac{35}{96 r^9} - \frac{29}{32 r^8} 
   + \frac{275}{192 r^7}\notag \\
& + \frac{11}{24 r^6} + \frac{185}{128 r^5} 
   - \frac{145}{192 r^4} - \frac{25}{96 r^3}.
\end{align}
For $l=0$, the integral \eqref{ISkregular} is expanded in $\alpha_0$ as
\begin{align}
	I[\mS^{(1)}](r)  &= -\frac{1}{4} + \frac{3}{4 r^3} - \frac{1}{4 r^2} - \frac{1}{4 r},\\[1mm]
	I[\mS^{(2)}](r) &= -\frac{3}{16} - \frac{3}{16 r^6} + \frac{3}{16 r^5} + \frac{1}{8 r^4} + \frac{5}{8 r^3} - \frac{5}{16 r^2}\notag\\
	&\quad - \frac{1}{4 r},\\[1mm]
	I[\mS^{(3)}](r) &= -\frac{1}{6} + \frac{3}{128 r^9} - \frac{1}{384 r^8} + \frac{5}{96 r^7} - \frac{29}{192 r^6} + \frac{55}{192 r^5}\notag\\
	&\quad + \frac{11}{96 r^4}  + \frac{185}{384 r^3} - \frac{145}{384 r^2} - \frac{25}{96 r}.
\end{align}
From the above results, it can be seen that, similar to the Bardeen BH, 
the case with $l=0$ exhibits non-vanishing TLN, and no logarithmic term appears. The first-, second- and third-order corrections all contribute negatively to the TLNs. 
Since we have chosen the parameters $x = 1$ and $c = 3$, this type of BH shows the same asymptotic behavior as the Hayward BH at large $r$. In this case, the scalar TLNs for $l=0$ can be written as 
\begin{equation}
	K_{l=0}^{0} 
	= -\frac{1}{4} \, \alpha_{0} 
	- \frac{3}{16} \, \alpha_{0}^{2} 
	- \frac{1}{6} \, \alpha_{0}^{3} 
	+ \mathcal{O}(\alpha_{0}^{4}) \,.
\end{equation}
For \(l=1\), the TLNs are computed up to third order of $\alpha_0$. The zeroth-order solutions~(\ref{eq:Phi_plus}) and~(\ref{eq:Phi_minus}) take the following form:
\begin{align}
\Phi^{(0)}_{+}(r)
&= \Phi^{(0)}(r)
= \Phi_{\mathrm{hor\text{-}reg}}(r)
= -\frac{r}{2} + r^2 , \\[2mm]
\Phi^{(0)}_{-}(r)
&= -12\, r + 6\, r \, \ln\!\left(1 - \frac{1}{r}\right)
   - 12\, r^2 \, \ln\!\left(1 - \frac{1}{r}\right) .
\end{align}
The TLNs can be obtained from the following expressions:
\begin{align}
	I[\mS^{(1)}](r)  &= -\frac{13}{48} + \frac{1}{16 r^3} - \frac{17}{48 r^2} + \frac{31}{48 r} + \frac{r}{12} - \frac{r^2}{6},\\[2mm]
	I[\mS^{(2)}](r) &= -\frac{35}{192} - \frac{1}{64 r^6} + \frac{7}{64 r^5} - \frac{19}{96 r^4} + \frac{11}{96 r^3}\notag\\
	&\quad - \frac{65}{192 r^2} + \frac{53}{96 r} + \frac{r}{12} - \frac{r^2}{8},\\[1mm]
	I[\mS^{(3)}](r)  &= -\frac{289}{2016} + \frac{1}{512 r^9} - \frac{313}{23040 r^8} + \frac{5459}{161280 r^7}\notag\\
	&\quad - \frac{1181}{23040 r^6} + \frac{3373}{23040 r^5} - \frac{6209}{23040 r^4} + \frac{3583}{23040 r^3}\notag\\
	&\quad - \frac{1591}{4608 r^2} + \frac{587}{1152 r} + \frac{25 r}{288} - \frac{r^2}{9}.
\end{align}
Here, we can see that for the scalar perturbations with $l=1$, the TLNs are nonzero and no logarithmic term appears. The contributions to the TLNs exhibit negative behavior 
similar to the case of \(l = 0\). This is in contrast to the Bardeen BH, whose TLN is positive and has a negative logarithmic running for $l=1$.
All contributions up to the third order can be accurately expressed as
\begin{equation}
	K_{l=1}^{0} = -\frac{13}{48} \, \alpha_{0} - \frac{35}{192} \, \alpha_{0}^{2} - \frac{289}{2016} \, \alpha_{0}^{3} + \mathcal{O}(\alpha_{0}^{4}) \,.
\end{equation}
\subsubsection{Vector field response}

Next, we compute the TLNs for vector perturbations ($s=1$). Similarly, for the case $l=1$, the TLNs can be obtained from the following integrations
\begin{align}
	I[\mS^{(1)}](r) &= -\frac{13}{12} + \frac{5}{4 r} - \frac{r}{12} - \frac{r^2}{12},\\[2mm]
	I[\mS^{(2)}](r) &= -\frac{35}{48} - \frac{7}{16 r^4} + \frac{7}{48 r^3} + \frac{1}{8 r^2} + \frac{25}{24 r} - \frac{r}{12}
	 - \frac{r^2}{16},\\[2mm]
	I[\mS^{(3)}](r)  &= -\frac{4687}{8064} + \frac{65}{896 r^7} - \frac{103}{1152 r^6} - \frac{317}{5760 r^5} - \frac{3359}{5760 r^4}\notag\\
	&\quad + \frac{1531}{5760 r^3} + \frac{209}{1152 r^2} + \frac{1073}{1152 r} - \frac{25 r}{288} - \frac{r^2}{18}.
\end{align}
Similar to scalar TLNs, we can clearly see that the TLNs are nonzero and that no logarithmic running appears and the contributions are also negative. The contributions to the TLN up to third order can be explicitly written as
\begin{align}
	K_{l=1}^{1} &= 
	-\frac{13}{12}\, \alpha_{0} 
	-\frac{35}{48}\, \alpha_{0}^2 
	-\frac{4687}{8064}\, \alpha_{0}^3 
	+ \mathcal{O}(\alpha_{0}^{4}) \,.
\end{align}
For \(l=2\), we compute the TLN only up to second order of $\alpha_{0}$, as the integral \eqref{Inhomogsolu} cannot be performed analytically, so we cannot construct the third-order source term. The explicit TLNs for the vector perturbations can be extracted from the following expressions:
	\begin{align}
		I[\mS^{(1)}](r)& = \frac{3}{320} + \frac{27}{64 r} - \frac{129 r}{320} - \frac{r^2}{320} + \frac{r^3}{10} - \frac{r^4}{8} \notag\\
        &+ \frac{9 \ln r}{10},\\
		I[\mS^{(2)}](r)  &= -\frac{32641}{256} + \frac{9 \pi^2}{80} - \frac{189}{1280 r^4} - \frac{1161}{1280 r^3} - \frac{1449}{128 r^2} \notag\\
        &+ \frac{21889}{160 r} + \frac{81 \pi^2}{16 r} - \frac{627 r}{640}- \frac{387 \pi^2 r}{80} - \frac{6341 r^2}{1280}\notag\\
        &- \frac{3 \pi^2 r^2}{80}
        + \frac{91 r^3}{10} + \frac{6 \pi^2 r^3}{5} - \frac{3 r^4}{32} - \frac{3 \pi^2 r^4}{2} \notag\\
		&+ \frac{2427 \ln r}{80} + \frac{108 \pi^2}{5} \ln r - \frac{9 \ln r}{16 r^3} - \frac{351 \ln r}{80 r^2} \notag\\
        &+ \frac{6333 \ln r}{80 r} - \frac{3 r \ln r}{8} - \frac{27 r^2 \ln r}{10} + 9 r^3 \ln r \notag\\
		& + \frac{27 \ln^2 r}{80} + \frac{243 \ln^2 r}{16 r} - \frac{1161 r \ln^2 r}{80}\notag\\
        &- \frac{9 r^2 \ln^2 r}{80} 
        + \frac{18 r^3 \ln^2 r}{5} - \frac{9 r^4 \ln^2 r}{2} + \frac{54 \ln^3 r}{5} \notag \notag\\
        &+ \frac{27}{40} \mathrm{Li}_2(1-r) 
        + \frac{243\, \mathrm{Li}_2(1-r)}{8 r} - \frac{1161 r\, \mathrm{Li}_2(1-r)}{40}\notag\\
        &- \frac{9 r^2\, \mathrm{Li}_2(1-r)}{40} 
		+ \frac{36 r^3\, \mathrm{Li}_2(1-r)}{5} - 9 r^4\, \mathrm{Li}_2(1-r) \notag\\
        &+ \frac{324 \ln r\, \mathrm{Li}_2(r)}{5}- \frac{648\, \mathrm{Li}_3(r)}{5} + \frac{648 \zeta(3)}{5}.
	\end{align}
Here, the explicit form of the first-order TLN is
$K_{l=1}^{1\,(1)} = {3}/{320} - {9 }/{10}\,\ln r$. The second-order TLN  is obtained from the asymptotic behavior of $I[\mS^{(2)}](r)$ at large $r$, which is given by
$K_{l=1}^{1\,(2)} = -{40025}/{256} + {648}/{5}\,\zeta(3) + {27}/{20} \ln r$. We observe that at first order $K_{l=1}^{1\,(1)}$ is positive while its running is negative. In contrast, at second order the coefficient $K_{l=1}^{1\,(2)}$ becomes negative, yet its running turns positive. Up to $\mathcal{O}(\alpha_{0}^{3})$, the TLN is given by
\begin{align}
K_{l=2}^{1} &= 
\left( \frac{3}{320} - \frac{9}{10}\, \ln r \right) \alpha_{0}\notag\\
        & + \left( -\frac{40025}{256} + \frac{648}{5}\, \zeta(3) + \frac{27}{20} \ln r \right) 
        \alpha_{0}^2\notag\\
        &+\mathcal{O}(\alpha_{0}^{3}). 
\end{align}

\subsubsection{Axial gravitational field response}
In this subsection, we compute the TLNs of the 
RBH with sub-Planckian curvature
under axial gravitational perturbations for $l=2$ and $l=3$. 
For the case of $l=2$, expanding the integral \eqref{ISkregular} in $\alpha_{0}$, the leading-order and next-to-leading-order terms are given by
\begin{align}
	I[\mS^{(1)}](r)  &= 
	\frac{337}{120} - \frac{51\,r}{20} - \frac{r^{2}}{10} 
	- \frac{r^{3}}{12} - \frac{3\,r^{4}}{40}, \\[4pt]
	I[\mS^{(2)}](r) &= 
	\frac{21491}{2880} - \frac{5729}{1920\,r^{3}} 
	- \frac{189}{160\,r^{2}} - \frac{3301}{5760\,r} 
	- \frac{803\,r}{320}\notag\\
	&\quad - \frac{7\,r^{2}}{80} 
	- \frac{17\,r^{3}}{240} - \frac{9\,r^{4}}{160} 
	+ \frac{2149}{2880}\,\ln r.
\end{align}

Similar to the case of the Bardeen BH, we find that 
the TLNs exhibit no logarithmic running at the first order, 
while a logarithmic dependence appears at the second order. And the contributions are both positive.
The TLN for $l=2$ up to the second order in $\alpha_{0}$ can be expressed as
\begin{align}
	K_{l=2}^{2} 
	= \frac{337}{120}\,\alpha_{0} 
	+ \left( \frac{21491}{2880} 
	+ \frac{2149}{2880}\,\ln r \right)\alpha_{0}^{2} 
	+ \mathcal{O}(\alpha_{0}^{3}).
\end{align}
For the case of $l=3$, expanding the integral \eqref{ISkregular} in $\alpha_{0}$, the leading-order and second-order terms are given by
\begin{align}
	I[\mS^{(1)}](r) &= 
	\frac{1691}{4320} - \frac{425\,r}{336} 
	+ \frac{95\,r^{2}}{72} - \frac{1301\,r^{3}}{3024} 
	- \frac{r^{4}}{672}\notag\\
	&\quad  + \frac{13\,r^{5}}{140} 
	- \frac{3\,r^{6}}{28}, \\[4pt]
	I[\mS^{(2)}](r) &= 
	-\frac{1427983}{3628800} - \frac{28747}{69120\,r^{3}} 
	+ \frac{51497}{50400\,r^{2}} + \frac{3562691}{7257600\,r} \notag\\
	&\quad 
	- \frac{867421\,r}{403200} + \frac{74633\,r^{2}}{40320} 
	- \frac{25831\,r^{3}}{60480} + \frac{13\,r^{4}}{13440} \notag\\
	&\quad 
	+ \frac{29\,r^{5}}{280} - \frac{9\,r^{6}}{112} 
	+ \frac{669067}{518400}\,\ln r.
\end{align}
Similar to the $l=2$ case, however, we note that the second-order 
constant term takes a negative value, indicating a weaker tidal 
response in comparison to the $l=2$ mode. The expansion of the 
TLN for $l=3$  up to the second order is given by
\begin{align}
	K_{l=3}^{2} 
	= &\frac{1691}{4320}\,\alpha_{0} 
	+ \left( 
	-\frac{1427983}{3628800} 
	+ \frac{669067}{518400}\,\ln r 
	\right)\alpha_{0}^{2} \notag\\
	&+ \mathcal{O}(\alpha_{0}^{3}).
\end{align}
It is worth noting that in Ref.~\cite{Barura:2024uog} the authors realized the Hayward metric as a background solution within the effective field theory (EFT) framework of scalar-tensor theories with a timelike scalar profile~\cite{Mukohyama:2023xyf} and evaluated the TLNs associated with axial gravitational perturbations.
In contrast, in the present work the deviations from the classical singular Schwarzschild BH are modeled entirely within classical GR by introducing a nontrivial stress-energy tensor corresponding to an anisotropic fluid. The RBH possessing sub-Planckian curvature for the parameter choice $x=1$ and $c=3$ exhibits the same asymptotic behavior as the Hayward solution. A remarkable feature emerging from our results is that the logarithmic running of the TLNs already appears at the lowest multipole $l=2$, whereas in the EFT-based realization it only shows up for $l\geq4$. This highlights that the TLNs are sensitive probes of the internal structure of RBHs and are capable of distinguishing between different RBH realizations.

\subsection{The regular black hole in asymptotically safe gravity}

In this subsection, we calculate the TLNs of the RBH \cite{Bonanno:2023rzk} in ASG under scalar, vector, and axial gravitational perturbations. Similarly, setting $r_h=1$ we can express the BH mass $M$ as a series expansion in terms of the deviation parameter $\xi$,
\begin{equation}
	M(\xi) = \frac{1}{2} + \frac{3\xi}{4} + \frac{3\xi^{2}}{4} + \frac{9\xi^{3}}{16} + \mathcal{O}(\xi^{4}) .
\end{equation}
Up to the same order in the parameter $\xi$, the metric function can be expanded as
\begin{align}
	f(r) &= 1 - \frac{1}{r} 
	- \frac{3(-1 + r^{3})\,\xi}{2 r^{4}}
	- \frac{3(-2 + r^{3})(-1 + r^{3})\,\xi^{2}}{2 r^{7}}\notag\\
	&\quad
	- \frac{9(-6 + 12 r^{3} - 7 r^{6} + r^{9})\,\xi^{3}}{8 r^{10}}
	+ \mathcal{O}(\xi^{4}) .
\end{align}
The effective potential for scalar field $(s = 0)$ 
and vector field $(s = 1)$ can be expanded in the same way as
	\begin{align}
		U_{1}(r) & = 
		\frac{3\,l(l + 1) - 3\,s^{2}}{2\,r^{3}}
		+ \frac{-3 + 6\,l(l + 1) - 6\,s^{2}}{4\,r^{4}} \notag\\
		&+ \frac{-3 + 3\,l(l + 1) - 3\,s^{2}}{2\,r^{5}}
		+ \frac{9\,(3 + 2\,s^{2})}{4\,r^{6}} , \\[4pt]
		U_{2}(r) & = 
		\frac{3\,l(l + 1) - 3\,s^{2}}{2\,r^{3}}
		+ \frac{-21 + 60\,l(l + 1) - 60\,s^{2}}{16\,r^{4}} \notag\\
		&+ \frac{-69 + 96\,l(l + 1) - 96s^{2}}{16\,r^{5}} 
		+ \frac{72 + 15l(l + 1) + 39s^{2}}{4\,r^{6}}\notag\\
        &+ \frac{93 + 12l(l + 1) + 42s^{2}}{8\,r^{7}}+ \frac{273 - 12l(l + 1) + 120s^{2}}{16\,r^{8}}\notag\\
        &- \frac{9\,(69 + 20\,s^{2})}{16\,r^{9}} , \\[4pt]
		U_{3}(r) &= 
		\frac{9\,(2l + 2l^{2} - 2s^{2})}{16\,r^{3}}
		+ \frac{9\,(-3 + 10l + 10l^{2} - 10s^{2})}{16\,r^{4}} \notag\\
		&- \frac{9(15 - 24l-24l^{2} + 24s^{2})}{16\,r^{5}} 
		+ \frac{9\,(42 + 30l + 30l^{2} + 12s^{2})}{16\,r^{6}}\notag\\
		&+ \frac{9\,(66 + 24l + 24l^{2} + 24s^{2})}{16\,r^{7}}
		+ \frac{9\,(141 + 6l + 6l^{2} + 60s^{2})}{16\,r^{8}} \notag\\
		&
		+ \frac{9\,(-243 - 60s^{2})}{16\,r^{9}}
		+ \frac{9\,(-82 - 2l - 2l^{2} - 16s^{2})}{16\,r^{10}} \notag\\
		&+ \frac{9\,(-152 + 2l + 2l^{2} - 38s^{2})}{16\,r^{11}}
		+ \frac{9\,(288 + 54s^{2})}{16\,r^{12}} .
	\end{align}
In addition, the first-, second-, and third-order effective potential contributions for axial gravitational perturbations can be written explicitly as follows:
	\begin{align}
		U_{1}(r) &= 
		\frac{-12 + 3 l (l + 1)}{2 r^{3}}
		+ \frac{-27 + 6 l (l + 1)}{4 r^{4}}\notag\\
		&+ \frac{-15 + 3 l (l + 1)}{2 r^{5}}
		+ \frac{171}{4 r^{6}}, \\[4pt]
		U_{2}(r) &=
		\frac{-12 + 3 l (l + 1)}{2 r^{3}}
		+ \frac{-261 + 60 l (l + 1)}{16 r^{4}}\notag\\
		&+ \frac{-453 + 96 l (l + 1)}{16 r^{5}}
		+ \frac{444 + 15 l (l + 1)}{4 r^{6}} \notag\\
		&+ \frac{477 + 12 l (l + 1)}{8 r^{7}}
		+ \frac{1185 - 12 l (l + 1)}{16 r^{8}}\notag\\
		&- \frac{2925}{16 r^{9}}, \\[4pt]
		U_{3}(r) &=
		\frac{-36 + 9 l (l + 1)}{8 r^{3}}
		+ \frac{-387 + 90 l (l + 1)}{16 r^{4}}\notag\\
		&+ \frac{27 (-37 + 8 l (l + 1))}{16 r^{5}}
		+ \frac{27 (43 + 5 l (l + 1))}{8 r^{6}} \notag\\
		&+ \frac{27 (59 + 4 l (l + 1))}{8 r^{7}}+ \frac{9 (645 + 6 l (l + 1))}{16 r^{8}}\notag\\
		&- \frac{10395}{16 r^{9}}
		- \frac{9 (181 + l (l + 1))}{8 r^{10}} \notag\\
		&+ \frac{9 (-296 + l (l + 1))}{8 r^{11}}
		+ \frac{1377}{2 r^{12}} .
	\end{align}
Next, we compute the TLNs of the RBH in ASG for scalar, vector, and axial gravitational perturbations.

\subsubsection{Scalar field response}
We first compute the TLNs of the RBH in ASG under scalar perturbations for \(l=0\) and $l=1$. 
For the case \(l=0\), the zeroth-order (\(\mathcal{O}(\xi^{0})\) modes, representing the growing and decaying solutions in Eqs.~(\ref{eq:Phi_plus}) and (\ref{eq:Phi_minus}), take the explicit forms
\begin{align}
\Phi^{(0)}_{+}(r)
&= \Phi^{(0)}(r)
= \Phi_{\mathrm{hor\text{-}reg}}(r)
= r , \\[2mm]
\Phi^{(0)}_{-}(r)
&= -\, r \, \ln\!\left(1-\frac{1}{r}\right) .
\end{align}

When \(s = 0\) and \(l = 0\), the first-order source term at \(\mathcal{O}(\xi^{1})\) can be evaluated from Eq.~(\ref{SKr}), yielding
\begin{align}
S^{(1)}(r) = U^{(1)}(r)\, \Phi^{(0)}(r)
= \frac{27}{4 r^5} - \frac{3}{2 r^4} - \frac{3}{4 r^3}.
\end{align}

Using the Green’s function approach, the first-order solution and the corresponding second-order source term \(\mathcal{O}(\xi^{2})\) are obtained as
\begin{align}
\Phi^{(1)}(r)
&= \int_{1}^{\infty} G(r,r')\, S^{(1)}(r')\, dr'
= -\frac{3}{4} - \frac{3}{4 r^2} - \frac{3}{4 r}, \\[2mm]
S^{(2)}(r)
&= U^{(2)}(r)\,\Phi^{(0)}(r) + U^{(1)}(r)\,\Phi^{(1)}(r) \notag \\
&= -\frac{351}{8 r^8} + \frac{105}{8 r^7} + \frac{33}{4 r^6} + \frac{315}{16 r^5} - \frac{15}{4 r^4} - \frac{21}{16 r^3}.
\end{align}

Proceeding to the next order, the second-order solution and the third-order source term at \(\mathcal{O}(\xi^{3})\), as indicated by Eqs.~(\ref{Inhomogsolu}) and (\ref{SKr}), are given by
\begin{align}
\Phi^{(2)}(r)
&= \int_{1}^{\infty} G(r,r')\, S^{(2)}(r')\, dr' \notag \\
&= -\frac{3}{4} + \frac{39}{32 r^5} + \frac{15}{16 r^4} + \frac{21}{32 r^3} - \frac{21}{16 r^2} - \frac{33}{32 r}, \\[2mm]
S^{(3)}(r)
&= U^{(3)}(r)\, \Phi^{(0)}(r)
   + U^{(2)}(r)\, \Phi^{(1)}(r)
   + U^{(1)}(r)\, \Phi^{(2)}(r) \notag \\
&= \frac{25515}{128 r^{11}} - \frac{1035}{16 r^{10}} - \frac{2331}{64 r^9} - \frac{729}{4 r^8} 
   + \frac{1755}{32 r^7}\notag \\
& + \frac{405}{16 r^6} + \frac{3807}{128 r^5} - \frac{441}{64 r^4} - \frac{27}{16 r^3}.
\end{align}

For \(l=0\), we can extract the TLNs from the following functions:
\begin{align}
	I[\mS^{(1)}](r) &= -\frac{3}{4} + \frac{9}{4\,r^3} - \frac{3}{4\,r^2} - \frac{3}{4\,r}, \\[2mm]
	I[\mS^{(2)}](r) &= -\frac{3}{4} - \frac{117}{16\,r^6} + \frac{21}{8\,r^5} + \frac{33}{16\,r^4} + \frac{105}{16\,r^3} - \frac{15}{8\,r^2}\notag\\
	&\quad - \frac{21}{16\,r}, \\[2mm]
	I[\mS^{(3)}](r) &= -\frac{9}{16} + \frac{2835}{128\,r^9} - \frac{1035}{128\,r^8} - \frac{333}{64\,r^7} - \frac{243}{8\,r^6}\notag\\ &+ \frac{351}{32\,r^5} 
	+ \frac{405}{64\,r^4} + \frac{1269}{128\,r^3} - \frac{441}{128\,r^2} - \frac{27}{16\,r}.
\end{align}
From the above, we can see that the TLNs are nonzero and no logarithmic running appears. Moreover, the contribution at each order of $\xi$ to TLN is all negative, showing similar behavior to  the RBH with sub-Planckian curvature. Thus, up to the third order of $\xi$  the scalar TLN for \(l=0\) is expressed as
\begin{equation}
	K_{l=0}^{0} = 
	-\frac{3}{4}\,\xi 
	-\frac{3}{4}\,\xi^2 
	-\frac{9}{16}\,\xi^3 
	+ \mathcal{O}(\xi^4).
\end{equation}
For $l=1$, the scalar TLNs can be obtained from the following expressions:
	\begin{align}
		I[\mS^{(1)}](r) &= -\frac{13}{16} + \frac{3}{16\, r^3} - \frac{17}{16\, r^2} + \frac{31}{16\, r} + \frac{r}{4} - \frac{r^2}{2}, \\[2mm]
		I[\mS^{(2)}](r) &= -\frac{35}{32} - \frac{39}{64\, r^6} + \frac{13}{4\, r^5} - \frac{299}{64\, r^4} + \frac{71}{64\, r^3} - \frac{95}{32\, r^2}\notag\\
        &+ \frac{323}{64\, r} + \frac{7\, r}{16} - \frac{r^2}{2}, \\[2mm]
		I[\mS^{(3)}](r) &= -\frac{591}{448} + \frac{945}{512 \, r^9} - \frac{24609}{2560 \, r^8} 
		+ \frac{7611}{560 \, r^7}\notag\\
        &- \frac{6399}{1280 \, r^6} + \frac{17247}{1280 \, r^5} - \frac{3237}{160 \, r^4} + \frac{9099}{2560 \, r^3} \notag\\
		&\quad - \frac{2475}{512 \, r^2} + \frac{267}{32 \, r} + \frac{9 \, r}{16} - \frac{3 \, r^2}{8}.
	\end{align}
Similar to the $l=1$ case, we can see that the TLNs at each order of $\xi$ are all negative and do not exhibit running terms.  The explicit expansion of the scalar TLN for $l=1$ can be written as
\begin{equation}
	K_{l=1}^{0} = 
	-\frac{13}{16}\, \xi 
	-\frac{35}{32}\, \xi^2 
	-\frac{591}{448}\, \xi^3 
	+ \mathcal{O}(\xi^4).
\end{equation}

\subsubsection{Vector field response}
Next, we compute the TLNs of the RBH in ASG under vector perturbations, for \(l=1\) and \(l=2\). 
For \(l=1\), the TLNs can be obtained from the following expressions:
\begin{align}
	I[\mS^{(1)}](r)  &= -\frac{13}{4} + \frac{15}{4\, r} - \frac{r}{4} - \frac{r^2}{4}, \\
	I[\mS^{(2)}](r)  &= -\frac{35}{8} - \frac{123}{16\, r^4} + \frac{13}{8\, r^3} + \frac{23}{16\, r^2} + \frac{155}{16\, r} - \frac{7 r}{16}\notag\\
	&\quad - \frac{r^2}{4}, \\
	I[\mS^{(3)}](r)  &= -\frac{4917}{896} + \frac{18063}{896\, r^7} - \frac{801}{128\, r^6} - \frac{93}{20\, r^5} - \frac{10221}{320\, r^4}\notag\\
	&\quad + \frac{2439}{320\, r^3} + \frac{45}{8\, r^2} + \frac{2007}{128\, r} - \frac{9 r}{16} - \frac{3 r^2}{16}.
\end{align}
Similar to scalar TLNs,  the  contributions to the TLN at each order are all negative and no logarithmic running appears. Up to the third order, the explicit expansion of the vector TLN for $l=2$ can thus be written as
\begin{equation}
	K_{l=1}^{1} = 
	-\frac{13}{4}\, \xi 
	-\frac{35}{8}\, \xi^2 
	-\frac{4917}{896}\, \xi^3
	+ \mathcal{O}(\xi^4).
\end{equation}
For \(l=2\), we mainly compute the first two orders of the TLNs. 
As mentioned earlier, the third-order integral \eqref{ISkregular} cannot be obtained analytically. 
The TLN can be obtained from the following expressions:
	\begin{align}
		I[\mS^{(1)}](r) &= \frac{9}{320} + \frac{81}{64\, r} - \frac{387\, r}{320} - \frac{3\, r^2}{320} + \frac{3\, r^3}{10} - \frac{3\, r^4}{8}\notag\\
		&+ \frac{27 \ln r}{10}, \\[2mm]
		I[\mS^{(2)}](r) &= -\frac{73383}{64} + \frac{81 \pi^2}{80} 
		- \frac{3321}{1280 r^4} - \frac{2457}{640 r^3} - \frac{27135}{256 r^2} \notag\\
        &+ \frac{1576023}{1280 r} + \frac{729 \pi^2}{16 r}   - \frac{11271 r}{1280} - \frac{3483 \pi^2 r}{80}\notag\\
        &- \frac{28527 r^2}{640}
		- \frac{27 \pi^2 r^2}{80} 
		+ \frac{3261 r^3}{40} + \frac{54 \pi^2 r^3}{5}\notag\\
        &- \frac{3 r^4}{8} - \frac{27 \pi^2 r^4}{2} 
        + \frac{21843 \ln r}{80} + \frac{972 \pi^2 \ln r}{5} \notag\\
        &- \frac{81 \ln r}{16 r^3} - \frac{3159 \ln r}{80 r^2}
		+ \frac{56997 \ln r}{80 r}\notag\\
        &- \frac{27 r \ln r}{8}- \frac{243 r^2 \ln r}{10} + 81 r^3 \ln r \notag\\
        &+ \frac{243 \ln^2 r}{80} + \frac{2187 \ln^2 r}{16 r} - \frac{10449 r \ln^2 r}{80}\notag\\
		& - \frac{81 r^2 \ln^2 r}{80} 
		+ \frac{162 r^3 \ln^2 r}{5} - \frac{81 r^4 \ln^2 r}{2} \notag\\
        &+ \frac{486 \ln^3 r}{5} + \frac{243}{40} \mathrm{Li}_2(1-r) + \frac{2187\, \mathrm{Li}_2(1-r)}{8 r}\notag\\
		& - \frac{10449 r\, \mathrm{Li}_2(1-r)}{40}
		- \frac{81 r^2\, \mathrm{Li}_2(1-r)}{40} \notag\\
        &+ \frac{324 r^3\, \mathrm{Li}_2(1-r)}{5} - 81 r^4\, \mathrm{Li}_2(1-r)  \notag\\
        &+ \frac{2916 \ln r\, \mathrm{Li}_2(r)}{5}- \frac{5832}{5} \mathrm{Li}_3(r) + \frac{5832 \zeta(3)}{5}.
	\end{align}
We find that the first-order contribution to the TLN is positive, while the logarithmic running behaves inversely. The second-order correction of the vector TLN can be obtained from the asymptotic behavior of $I[\mS^{(2)}](r)$ at large $r$. Up to $\mathcal{O}(\xi^{3})$, the vector TLN for $l=2$ is given by
\begin{align}
K_{l=2}^{1} &= 
\left( \frac{9}{320} - \frac{27}{10} \ln r \right) \xi\notag\\
        &+ \left( -\frac{89997}{64} + \frac{5832}{5}\, \zeta(3) + \frac{243}{20} \ln r \right) \xi^2+ \mathcal{O}(\xi^3),
\end{align}
from which one can see that the second-order contribution is negative while the running is positive.

\subsubsection{Axial gravitational field response}

Finally, we compute the TLNs of the RBH under axial gravitational perturbations within the framework of 
ASG for $l=2$ and $l=3$. 
For the case of $l=2$, the expansion of the integral \eqref{ISkregular} in $\xi$ is given by
\begin{align}
	I[\mS^{(1)}](r)  &= 9 - \frac{171 r}{20} - \frac{3 r^{2}}{20} - \frac{3 r^{3}}{20} - \frac{3 r^{4}}{20}, \\[4pt]
	I[\mS^{(2)}](r) &= \frac{159}{2} - \frac{513}{16 r^{3}} - \frac{927}{40 r^{2}} 
	+ \frac{177}{80 r} - \frac{411 r}{16} - \frac{3 r^{2}}{8}\notag\\
	&\quad
	- \frac{21 r^{3}}{80} - \frac{3 r^{4}}{20}, \\[4pt]
	I[\mS^{(3)}](r) &= \frac{269379}{11200} + \frac{45333}{448 r^{6}}
	+ \frac{1832517}{22400 r^{5}} + \frac{1538937}{22400 r^{4}}\notag\\
	&\quad
	- \frac{1060911}{5600 r^{3}} - \frac{63063}{800 r^{2}}  + \frac{62289}{1600 r} - \frac{28971 r}{640}\notag\\
	&\quad
	- \frac{441 r^{2}}{640} - \frac{27 r^{3}}{80} - \frac{9 r^{4}}{80}.
\end{align}
Consequently, the TLN for $l=2$ up to the third order can directly read as
\begin{equation}
	K_{l=2}^{2} 
	= 9\,\xi
	+ \frac{159}{2}\,\xi^{2}
	+ \frac{269379}{11200}\,\xi^{3}
	+ \mathcal{O}(\xi^{4}) .
\end{equation}
It is evident that the TLN is nonzero in this case. 
However, we find that, even when computed up to the third order, 
no logarithmic running term appears for the $l=2$ mode under 
axial gravitational perturbations in ASG, in contrast to the previous two models. Moreover, their contributions to the TLN are all positive, which is consistent with the previous two types of RBHs.

For the case of $l=3$, the expansion of the integral \eqref{ISkregular} in $\xi$ is given as follows:
\begin{align}
	I[\mS^{(1)}](r) &= 
	\frac{9}{7} - \frac{475 r}{112} + \frac{1535 r^{2}}{336}
	- \frac{529 r^{3}}{336} - \frac{r^{4}}{336}
	+ \frac{r^{5}}{4}\notag\\
	&\quad - \frac{2 r^{6}}{7}, \\[4pt]
	I[\mS^{(2)}](r) &= 
	-\frac{208181}{6720} - \frac{513}{112 r^{3}} + \frac{10763}{2240 r^{2}}
	+ \frac{124613}{3360 r}\notag\\
	&\quad - \frac{19157 r}{960}
	+ \frac{1529 r^{2}}{84} - \frac{907 r^{3}}{192}  - \frac{5 r^{4}}{672} + \frac{7 r^{5}}{16}\notag\\
	&\quad - \frac{2 r^{6}}{7}
	+ \frac{237 \ln r}{7}, \\[4pt]
	I[\mS^{(3)}](r) &= 
	-\frac{143685019}{376320} - \frac{120825}{12544 r^{6}}
	- \frac{1309101}{31360 r^{5}}\notag\\
	&\quad - \frac{4498499}{125440 r^{4}}
	+ \frac{3584485}{18816 r^{3}}  - \frac{8035661}{125440 r^{2}}\notag\\
	&\quad
	+ \frac{2444089}{8960 r} + \frac{57889 r}{1280}
	+ \frac{114045 r^{2}}{3584}\notag\\
	&\quad - \frac{19847 r^{3}}{2688}
	+ \frac{57 r^{4}}{112}  - \frac{69 r^{5}}{560} - \frac{3 r^{6}}{14}\notag\\
	&\quad
	+ \frac{320991 \ln r}{4480}.
\end{align}
Unlike the $l=2$ case, here the logarithmic running terms appear 
at both the second and third orders. Only the first-order term contributes positively to the TLNs, while the second- and third-order terms contribute negatively.
Accordingly, the TLNs up to the third order can be expressed as
\begin{align}
	K_{l=3}^{2}& 
	= \frac{9}{7}\,\xi
	+ \left( -\frac{208181}{6720} + \frac{237 \ln r}{7} \right)\xi^{2}\notag\\
	&\quad
	+ \left( -\frac{143685019}{376320} + \frac{320991 \ln r}{4480} \right)\xi^{3}
	+ \mathcal{O}(\xi^{4}).
\end{align}
This indicates that for $l=3$, the logarithmic contributions emerge beyond 
the leading order, marking a qualitative difference from the $l=2$ case 
in the axial gravitational perturbations of the RBH in ASG.

\subsection{Renormalization-group interpretation of the tidal response}

In this subsection, we clarify the RG interpretation of the logarithmic behavior appearing in the tidal response, following the general logic of the EFT treatments in Refs.~\cite{Combaluzier--Szteinsznaider:2025eoc,Caron-Huot:2025tlq}.

In the asymptotic region, the perturbation equation \eqref{PerturbationEqModi} admits solution of the schematic form
\begin{equation}
\Phi(r) \sim r^{\ell+1}
+ \left[ K_\ell + \beta_\ell \ln r \right] r^{-\ell} ,
\label{eq:asymptotic_log_structure}
\end{equation}
where $K_\ell$ is the tidal Love number and $\beta_\ell$ is the coefficient of the logarithmic term. 
The presence of the logarithm implies that the tidal response depends on the radial scale at which it is defined. This is the classical analog of renormalization-group running.

To make this scale dependence explicit, we introduce a renormalized Love number defined at a reference scale $\mu$,
\begin{equation}
K_\ell(\mu)
= K_\ell(r) + \beta_\ell \ln(\mu r).
\label{eq:renormalized_TLN}
\end{equation}
Requiring that physical observables be independent of the arbitrary scale $\mu$ leads to the RG equation
\begin{equation}
\mu \frac{d K_\ell}{d\mu}
= -\beta_\ell .
\label{eq:RG_equation}
\end{equation}

The scale $\mu$ has a clear physical interpretation. 
In the asymptotic solution, the logarithmic term indicates that the induced multipole moment depends on the distance at which the tidal response is probed. 
The renormalization procedure amounts to defining the Love number at a reference radius $r_\ast$, with
\begin{equation}
\mu \sim r_\ast^{-1}.
\end{equation}
This scale separates the near-zone physics associated with the internal structure of the BH from the far-zone region where the tidal field is measured. 
Different choices of $\mu$ correspond to different conventions for defining the Love number, while physical observables remain invariant once the RG running is taken into account.

This structure is consistent with the EFT-based analyses of Refs.~\cite{Combaluzier--Szteinsznaider:2025eoc,Caron-Huot:2025tlq}, where the scale $\mu$ arises as the renormalization scale introduced in dimensional regularization. 
In contrast to the dynamical EFT treatments, where the running originates from ultraviolet divergences in worldline loop diagrams, in the present static problem the RG structure emerges directly from the logarithmic behavior of the asymptotic perturbation solutions.

\subsection{Comparison of three regular black holes }
In Table~\ref{tab:tlns_summary}, we present the leading-order tidal Love numbers 
(\(C_{\mathrm{const}}, C_{\ln}\)) for RBHs
under scalar, vector, and axial gravitational perturbations. 
All regular parameters are evaluated at \(q^2 = 0.1\), \(\alpha_0 = 0.1\), 
and \(\xi = 0.1\) for the Bardeen BH, the BH with sub-Planckian curvature and the RBH in ASG, respectively. 
We can clearly observe both the connections and distinctions among the three models. 
The key finding is that all RBH models exhibit generally nonzero TLNs under scalar, vector, 
and gravitational perturbations, signaling a breakdown of the classical ``no-hair'' theorem 
in the quantum gravity regime and confirming a dynamical response of the internal structure 
to external tidal fields.

From Table ~\ref{tab:tlns_summary} we can identify several qualitative discriminants among the three RBH models, which may prove useful in future phenomenological assessments of their distinguishability. \\
{(1) \em Scalar sector.} For the monopole mode $l=0$, all models exhibit negative, scale-independent TLNs, so this mode is not very diagnostic. However, the dipole mode $l=1$ shows a clear separation: the Bardeen BH yields a positive TLN with non-vanishing logarithmic running, whereas both the sub-Planckian and ASG BHs give negative TLNs without any running. Thus, the sign and scale dependence of the scalar dipole TLN already distinguish de Sitter-core geometries from Minkowski-core and ASG BHs.\\
{(2) \em Vector sector.} In the dipole mode $l=1$, the Bardeen BH again stands out by having a positive TLN accompanied by a sizable negative logarithmic term, while the other two models have negative, non-running TLNs. For the quadrupole mode $l=2$, the three models occupy distinct regions in the  (\(C_{\text{const}}\), \(C_{\text{log}}\)): Bardeen has  (\(C_{\text{const}}<0\), \(C_{\text{log}}<0\)),  the sub-Planckian BH has a very small positive constant and a moderate positive running, and the ASG BH shows a slightly positive constant but a much larger positive running. This pattern provides a two-dimensional “fingerprint’’ for discriminating the three scenarios.\\
{(3)\em Axial gravitational sector.} The axial quadrupole TLN $l=2$, which is expected to be the most relevant for GW observations, exhibits a pronounced enhancement in the ASG case: At the same deformation parameters, the ASG BH has \(C_{\text{const}}\approx0.9\), about a factor of 3 larger than the Bardeen and sub-Planckian values $\sim 0.26-0.28$, while remaining free of logarithmic running. The octupole mode $l=3$ shows a similar hierarchy of magnitudes. These features indicate that the ASG BH leaves a much stronger imprint in the axial gravitational channel, offering a particularly clean theoretical signature of this quantum-gravity-motivated model.

These results indicate that TLNs not only serve as a useful criterion 
for distinguishing classical BHs from their quantum-corrected counterparts and provide a comparative theoretical window into the internal structure of RBHs and the underlying mechanisms that support them. They also motivate future phenomenological studies aimed at assessing how such tidal signatures may be tested with GW observations. In this paper, “distinguishability” refers to the presence of model-dependent theoretical tidal patterns—such as sign changes, logarithmic running, and sector-dependent hierarchies—rather than to a demonstrated observational resolvability in realistic detector data.

\subsection{Astrophysical relevance and outlook}
\label{sec:astro}

A central question is whether existing observations already force the deviation parameters of RBHs to be so small that the associated TLNs become negligible, and whether future GW detectors can improve this situation. In the models studied here, the deviations from Schwarzschild are controlled by $(q,\alpha_0,\xi)$ and arise primarily in the strong-field region. This can be seen directly from the large-$r$ expansions of the metric functions, which start at relatively high post-Newtonian (PN) order:
\begin{align}
f_{\rm Bardeen}(r) &= 1-\frac{2M}{r} + \frac{3M q^2}{r^3} + \mathcal{O}(r^{-5}),
\\
f_{\rm core}(r) &= 1-\frac{2M}{r} + \mathcal{O}(r^{-4}),
\\
f_{\rm ASG}(r) &= 1-\frac{2M}{r} + \frac{6M^2 \xi}{r^4} + \mathcal{O}(r^{-5}),
\end{align}
where $f_{\rm core}$ denotes the Minkowski-type core (sub-Planckian curvature) geometry considered in this work. Therefore, all three models agree with Schwarzschild through the $1/r$ term, and the first non-Schwarzschild corrections enter only at $1/r^3$ or $1/r^4$.

\subsubsection{Why existing observations do not automatically force tiny deviations?}
\paragraph{Solar-System/PPN tests.}
Solar-System experiments strongly constrain the parametrized post-Newtonian (PPN) parameters (notably $\gamma$ and $\beta$) that are tied to the $1/r$ and $1/r^2$ structures of the metric potentials. However, since the metrics above deviate from Schwarzschild only at $1/r^3$ or higher, the leading PPN parameters remain unchanged, and the strongest Solar-System bounds (e.g.,\ Cassini constraints on $\gamma-1$ at the $\sim 10^{-5}$ level) do not directly translate into stringent constraints on $(q,\alpha_0,\xi)$ \cite{Bertotti:2003rm}. Any constraints would have to come from higher-order (beyond-standard) PN/PPN effects, for which current measurements are far less constraining.

\paragraph{Binary pulsars.}
Precision pulsar timing provides powerful strong-field tests of gravity within specified underlying theories (e.g.,\ by constraining dipolar radiation or strong-field modifications of the binding energy). In the present work, we adopt a phenomenological metric approach for static RBHs; without committing to a specific underlying field content, pulsar constraints cannot be straightforwardly mapped into tight bounds on $(q,\alpha_0,\xi)$. Moreover, typical pulsar orbital separations satisfy $M/r \ll 1$, so higher-PN corrections such as $(M/r)^3$ or $(M/r)^4$ are parametrically suppressed in the conservative dynamics \cite{Freire:2024adf}.

\paragraph{EHT black-hole imaging.}
EHT measurements constrain horizon-scale observables (shadow diameter/ring scale), but current uncertainties remain at the ${\cal O}(10\%)$ level. Parametrized analyses show that even ${\cal O}(1)$ deviations entering at 3PN order can induce only percent-level changes in the shadow diameter, leaving such higher-PN deformations weakly constrained at present \cite{Psaltis:2020ctj}. Since our RBH metrics deviate from Schwarzschild only at $1/r^3$ or higher, current EHT observations do not yet impose strong constraints on $(q,\alpha_0,\xi)$.

\paragraph{Current gravitational-wave constraints.}
In quasi-circular inspirals, conservative tidal effects enter the GW phasing at high PN order. Recent high-SNR BBH observations can already place direct bounds on an effective tidal deformability parameter; for instance, the analysis of GW250114 yields a 90\% upper limit $\tilde\Lambda<34.8$ in a convention where $\Lambda=(2/3)k_2$ (with compactness absorbed into the definition of $k_2$) \cite{Andres-Carcasona:2025bni}. These constraints are fully consistent with vanishing TLNs in vacuum GR, but they do not yet exclude small nonzero TLNs at the level predicted by moderate deviations in phenomenological RBH metrics.

\subsubsection{Prospects for next-generation GW detectors and EMRIs}
Next-generation ground-based detectors (CE/ET) are expected to significantly improve the sensitivity to high-PN finite-size effects. Fisher forecasts exploiting worldline EFT waveform phasing indicate that CE+ET can bound the mass-weighted tidal parameter at the level $\tilde\Lambda \sim \mathcal{O}(10)$ for binary compact objects, with further improvement possible through a population of events \cite{Shterenberg:2024tmo}.

Even stronger prospects arise for EMRIs observed by LISA. Due to the large number of inspiral cycles in band and the long observation time, EMRIs are expected to be highly sensitive to finite-size tidal effects of the central object. Existing analyses indicate that the tidal response parameter can be constrained down to the $\sim 10^{-3}$--$10^{-2}$ range, depending on the supermassive BH spin and source properties, with high-spin systems yielding the tightest bounds \cite{Piovano:2022ojl}.

\subsubsection{Translating TLN measurements into bounds on $(q,\alpha_0,\xi)$}
Our analytic results provide explicit leading-order relations between the quadrupolar TLNs and the deviation parameters. For the axial gravitational $l=2$ response, the leading contributions read schematically
\begin{align}
k_2^{\rm(Bardeen)} &\simeq \frac{103}{40}q^2 + \cdots, \\
k_2^{\rm(core)} &\simeq \frac{337}{120}\,\alpha_0 + \cdots, \\
k_2^{\rm(ASG)} &\simeq 9\,\xi + \cdots,
\end{align}
where the ellipses denote higher-order terms (including logarithmic running in two of the three cases, as discussed in previous subsections).
Therefore, an observational bound $|k_2|<k_2^{\rm max}$ can be immediately re-expressed as prospective constraints on the RBH parameters:
\begin{align}
q \lesssim \left(\frac{40}{103}k_2^{\rm max}\right)^{1/2},\quad
\alpha_0 \lesssim \frac{120}{337}k_2^{\rm max},\quad
\xi \lesssim \frac{1}{9}k_2^{\rm max}.
\end{align}
This mapping illustrates the main astrophysical motivation of the present work: Current observations do not generically force $(q,\alpha_0,\xi)$ to be tiny in these RBH metrics, while future GW observations---especially EMRIs---can potentially constrain these parameters through their tidal response. Here the mapping from $k_2$ to $(q,\alpha_0,\xi)$ is intended as an illustrative theoretical translation rather than a detector-specific forecast. The relative constraining power of different source classes depends on whether one considers dimensionless or dimensional tidal-response coefficients, as well as on the source mass scale and waveform modeling.

We emphasize, however, that the relative constraining power of TLN effects compared with other waveform observables, such as inspiral phasing or ringdown signatures, is generally source-dependent and requires dedicated waveform-level analyses beyond the scope of the present work. The purpose of our analysis is to establish the comparative theoretical behavior of TLNs in representative RBH models and to provide explicit mappings between tidal-response coefficients and the underlying deviation parameters. These results furnish a concrete starting point for future studies aimed at identifying which source classes and detector configurations may offer the strongest discriminatory power.

\begin{table*}[htbp]
	\centering
	\caption{
		Leading-order tidal Love numbers (\(C_{\text{const}}\), \(C_{\text{log}}\)) 
		for regular black holes under scalar, vector, and axial gravitational perturbations.
		All coefficients are evaluated at \(q^2 = 0.1\), \(\alpha_0 = 0.1\), and \(\xi = 0.1\)
		for the Bardeen, sub-Planckian, and ASG black holes, respectively.
	}
	\label{tab:tlns_summary}
	\vspace{4pt}
	\setlength{\tabcolsep}{10pt}
	\renewcommand{\arraystretch}{1.2}
	\begin{tabular}{
			cl
			S[table-format=-1.6]
			S[table-format=-1.6]
			S[table-format=-1.6]
		}
		\toprule
		\textbf{Perturbation type} & \textbf{Mode \(l\)} 
		& {\textbf{Bardeen BH (\(q^2=0.1\))}} 
		& {\textbf{Sub-Planckian BH (\(\alpha_0=0.1\))}} 
		& {\textbf{ASG BH (\(\xi=0.1\))}} \\
		\midrule
		
		\multirow{4}{*}{Scalar}
		& \( l = 0 \) 
		& {$C_{\text{const}}=-0.075$} 
		& {$C_{\text{const}}=-0.025$}
		& {$C_{\text{const}}=-0.075$} \\
		
		& 
		& {$C_{\text{log}}=0$} 
		& {$C_{\text{log}}=0$} 
		& {$C_{\text{log}}=0$} \\
		
		& \( l = 1 \)
		& {$C_{\text{const}}=0.094$}
		& {$C_{\text{const}}=-0.027$}
		& {$C_{\text{const}}=-0.081$} \\
		
		&
		& {$C_{\text{log}}=-0.05$}
		& {$C_{\text{log}}=0$}
		& {$C_{\text{log}}=0$} \\
		\midrule
		
		\multirow{4}{*}{Vector}
		& \( l = 1 \)
		& {$C_{\text{const}}=0.05$}
		& {$C_{\text{const}}=-0.11$}
		& {$C_{\text{const}}=-0.33$} \\
		
		&
		& {$C_{\text{log}}=-0.2$}
		& {$C_{\text{log}}=0$}
		& {$C_{\text{log}}=0$} \\
		
		& \( l = 2 \)
		& {$C_{\text{const}}=-0.096$}
		& {$C_{\text{const}}=0.00094$}
		& {$C_{\text{const}}=0.0028$} \\
		
		&
		& {$C_{\text{log}}=-0.068$}
		& {$C_{\text{log}}=0.09$}
		& {$C_{\text{log}}=0.27$} \\
		\midrule
		
		\multirow{4}{*}{Axial Grav.}
		& \( l = 2 \)
		& {$C_{\text{const}}=0.26$}
		& {$C_{\text{const}}=0.28$}
		& {$C_{\text{const}}=0.9$} \\
		
		&
		& {$C_{\text{log}}=0$}
		& {$C_{\text{log}}=0$}
		& {$C_{\text{log}}=0$} \\
		
		& \( l = 3 \)
		& {$C_{\text{const}}=0.017$}
		& {$C_{\text{const}}=0.039$}
		& {$C_{\text{const}}=0.13$} \\
		
		&
		& {$C_{\text{log}}=0$}
		& {$C_{\text{log}}=0$}
		& {$C_{\text{log}}=0$} \\
		\bottomrule
	\end{tabular}
\end{table*}

\section{conclusions}\label{sectionV}

This paper presents a systematic theoretical study of the tidal response of three important classes of RBH models---Bardeen BH, the BH with sub-Planckian curvature, and the BH in the ASG framework---under static tidal fields, with a focus on calculating and analyzing their TLNs. In GR, the TLNs of four-dimensional, asymptotically flat BHs vanish, indicating that they do not develop induced multipole moments under external tidal fields, exhibiting a unique ``rigidity.'' However, beyond classical GR---e.g., when quantum-gravity effects or matter fields are considered in RBH models---this zero property can be violated. Therefore, TLNs provide a useful theoretical probe of BH interior structure, possible deviations from classical general relativity, and quantum-gravity-inspired modifications. Even when inspiral phasing or ringdown observables ultimately provide tighter bounds on deviations from classical black-hole geometries, the TLNs studied here remain of independent interest as complementary diagnostics of the compact-object response.  

The core methodology of this work is an analytic Green's function technique developed in Ref~\cite{Barura:2024uog}. For consistent comparisons across different RBH models, all horizons were normalized to \( r_h = 1 \). The mass \( M \) and metric function \( f(r) \) of each BH were then expanded as series around their respective deviation or regular parameters (\( q \) for Bardeen, \( \alpha_0 \) for sub-Planckian, and \( \xi \) for ASG BH). By introducing a field redefinition, all corrections to GR are encapsulated in an effective potential \( U(r) \). The perturbation equations are then solved order by order: The zeroth-order solution corresponds to the Schwarzschild background and is expressed in terms of hypergeometric functions, while higher-order solutions are obtained systematically via the Green's function formalism to handle the inhomogeneous terms. The TLNs are extracted from the coefficients of the \( r^{-l} \) terms in the asymptotic expansion at spatial infinity. Notably, when logarithmic terms appear (e.g., \( \ln r \)), their coefficients are also interpreted as part of the TLNs and can be understood in a classical RG framework as a ``$\beta$ function''~\cite{Hui:2020xxx,Cardoso:2017cfl,Kol:2011vg}.  

The study yields rich and revealing results. First, for all three classes of RBHs, the TLNs are generically nonzero under scalar, vector, and axial gravitational perturbations, in stark contrast to classical GR BHs. This directly confirms that the nontrivial core structures of RBHs (e.g., the de Sitter core of Bardeen BH or the Minkowski core of sub-Planckian curvature BH) significantly affect their tidal response. Second, the TLNs strongly depend on the perturbation type and multipole \( l \). For instance, in scalar perturbations of Bardeen BH, the \( l=0 \) TLNs are negative and free of logarithms, whereas \( l=1 \) exhibits logarithmic structure. Third, a key finding is the ubiquitous presence of logarithmic running. In many cases, first-order corrections do not contain logarithms, but second- or third-order corrections produce TLNs of the form \( K \sim \text{constant} + \beta \ln r \).  We have shown that these logarithms require a renormalized definition of the Love numbers at a reference scale and lead to a classical RG running, which we derived explicitly. 

From an astrophysical perspective, we highlighted that current weak-field and intermediate-field observations do not generically impose stringent bounds on the deviation parameters of these RBHs. At the same time, the analytic relations derived here between the quadrupolar TLNs and the deviation parameters provide a useful framework for translating prospective tidal-response constraints into bounds on $(q,\alpha_0,\xi)$. In this sense, the present results offer a concrete comparative benchmark for future phenomenological studies of RBHs. Even when inspiral phasing or ringdown observables ultimately provide tighter bounds on deviations from classical BH geometries, the TLNs studied here remain of independent interest as complementary diagnostics of the compact-object response. 

Overall, the main value of the present work is to provide a coherent comparative benchmark for tidal deformability in several representative RBH  spacetimes. The analytic results obtained here identify model-dependent tidal signatures, clarify how these signatures depend on perturbation sector and deviation parameters, and establish a useful foundation for future phenomenological studies aimed at assessing to what extent tidal response may help distinguish different RBH models from one another and from their classical general-relativistic counterparts. Looking forward, this work opens multiple promising directions. Theoretically, it would be interesting to study more realistic rotating RBHs  or to study responses under dynamical (nonstatic) tidal fields. For higher-order computations, combining numerical and analytic techniques becomes necessary due to the complexity of integrals. A full assessment of observational discrimination among different RBH models would require waveform-level analyses that incorporate source class, detector sensitivity, parameter correlations, and the relative roles of tidal, inspiral, and ringdown observables. We leave such investigations to future work. Finally, understanding the microscopic origin of logarithmic terms and their potential role in BH thermodynamics and the information paradox will likely provide a frontier connection between gravity, quantum theory, and differential geometry. 

\begin{acknowledgments}
The work was in part supported by NSFC Grant No. 12205104 and the startup funding of South China University of Technology.
\end{acknowledgments}

	%

\end{document}